\documentclass[preprint,preprintnumbers,amsmath,amssymb]{revtex4}
\usepackage[utf8]{inputenc}
\usepackage[T1]{fontenc}
\usepackage{bm}
\usepackage{newtxmath,newtxtext}
\usepackage{graphicx}
\usepackage{textcomp}
\usepackage{float}
\usepackage{booktabs}
\usepackage{dcolumn,enumerate}
\usepackage{ragged2e}
\usepackage{hyperref}
\usepackage{multirow}
\hypersetup{colorlinks,citecolor=blue}
\hypersetup{colorlinks=true, linkcolor=blue, filecolor=blue, urlcolor=blue}

\usepackage{xcolor}
\usepackage{epsfig}
\usepackage{caption}
\usepackage{appendix}
\usepackage{subcaption}
\usepackage{graphicx} 

\begin{document}

\title{\textbf{Penrose Process Efficiency and Irreducible Mass in Rotating Einstein–Born–Infeld Black Holes with Nonlinear Electrodynamics}}

\author{\textsuperscript{1}Urooj Fatima}
\email{wafa4494@gmail.com}

\author{\textsuperscript{1}G. Abbas}
\email{ghulamabbas@iub.edu.pk}

\affiliation{\textsuperscript{1}Department of Mathematics, The Islamia University of Bahawalpur, Bahawalpur-63100, Pakistan}

\date{}

\begin{abstract}

We investigate the extraction of rotational energy from rotating Einstein--Born--Infeld (EBI) black holes, where nonlinear electrodynamics introduces a radius-dependent effective charge modifying the spacetime geometry. Focusing on neutral test particles in the equatorial plane, we derive analytic expressions for their kinematics and establish conditions for negative energy orbits essential to the Penrose process using the near-horizon limit and Wald inequality. We present a closed-form expression for maximal energy extraction efficiency as a function of spin, charge, and the Born--Infeld parameter \(\beta\). Our numerical survey reveals that increasing charge and nonlinear Born–Infeld effects generally reduce horizon radius and ergoregion size, suppressing energy extraction efficiency compared to Kerr and often Kerr–Newman black holes. However, at certain spins and \(\beta\), the EBI geometry can enhance efficiency beyond Kerr–Newman. We also compute the irreducible mass, showing how nonlinear electromagnetic dynamics reduce the horizon area and the associated entropy proxy. These results provide a unified picture linking nonlinear electrodynamics, horizon structure, and energy extraction efficiency across relevant parameters.
\end{abstract}

\maketitle
\section{Introduction}

Black hole physics provides profound insights into fundamental gravitational theory and high-energy astrophysical phenomena \cite{wald1984general,misner1973gravitation}. Among energy extraction mechanisms, such as the Penrose process \cite{penrose1969gravitational}, Blandford--Znajek \cite{blandford1977electromagnetic}, and superradiance \cite{brito2015superradiance}, understanding black hole energy extraction is vital for probing their internal structure and explaining astrophysical events like relativistic jets and gamma-ray bursts \cite{meier2012black,piran2004physics}.

Traditionally, charged black holes have been studied within linear electrodynamics frameworks, for example, the Reissner--Nordstr\"{o}m solutions. However, nonlinear electrodynamics, particularly Born--Infeld (BI) theory, introduces significant modifications in strong electromagnetic fields. Originally proposed to regularize the divergent self-energy of point charges \cite{BornInfeld1934}, BI electrodynamics later gained renewed relevance as the low-energy effective action in string theory on D-branes \cite{Fradkin1985,Leigh1989}. When coupled with Einstein gravity, EBI black holes exhibit distinctive horizon structures, causal properties, and thermodynamic behavior compared to RN black holes \cite{Fernando2003,Dey2004BornInfeldBH,breton2005smarr,Hendi2010}.

Numerous studies have examined EBI black hole geometry and thermodynamics. Fernando and Krug \cite{fernando2003charged} highlighted how BI nonlinearity affects horizon multiplicity, while Dey \cite{Dey2004BornInfeldBH} analyzed horizon structure and singularity regularization focusing on the BI parameter. Breton extended thermodynamic analyses to anti--de Sitter spacetimes \cite{breton2005smarr}, and Hendi et al.\ explored slowly rotating EBI black holes, detailing modifications to geometry and thermodynamics \cite{hendi2009rotating}. Despite this progress, the impact of nonlinear electrodynamics on energy extraction—particularly through the Penrose process in rotating EBI black holes—remains insufficiently investigated.

Rotating EBI black holes generalize Kerr--Newman solutions by incorporating BI nonlinearities that alter their geometry and physical behavior. Recent methods to generate rotating regular black holes avoid the Newman--Janis complexification, introducing physical and symmetry-based arguments that produce nonsingular, physically regular rotating solutions relevant to nonlinear electrodynamics,including nonlinear electrodynamics models \cite{azreg2014static, azreg2014regular}.
 Energy extraction studies have focused mostly on non-rotating EBI black holes, indicating that BI corrections suppress available extractable energy relative to RN cases, primarily due to weaker electromagnetic fields near the black hole core \cite{breton2016energy,Pereira2015}. However, numerical investigations of rotating EBI solutions and particle dynamics \cite{ChengWang2024} have yet to evaluate Penrose process efficiencies.
Energy extraction studies have focused mostly on non-rotating EBI black holes, indicating that BI corrections suppress available extractable energy relative to RN cases, primarily due to weaker electromagnetic fields near the black hole core \cite{breton2016energy,Pereira2015}. However, numerical investigations of rotating EBI solutions and particle dynamics \cite{ChengWang2024} have yet to evaluate Penrose process efficiencies. This work addresses that gap by computing maximum rotational energy extraction efficiencies for rotating EBI black holes, elucidating how BI nonlinearities reshape the ergosphere, affect negative energy states, and modify the irreducible mass.

Given the unique properties of extremal EBI black holes—including the absence of a smooth Maxwell limit and distinctive thermodynamic features \cite{breton2005smarr,breton2016energy}—investigating nonlinear effects on energy extraction is critical. Our comprehensive analysis integrates detailed numerical tables with extensive graphical illustrations of event and Cauchy horizons, ergoregion deformation, shrinking, and splitting. This unified approach offers reproducible insights into how nonlinear electrodynamics and rotation jointly influence black hole causal structure and energy extraction efficiency, extending beyond the primarily qualitative or limited-parameter studies in existing literature.

This paper systematically investigates the properties of rotating EBI black holes. Section \ref{S2} introduces the rotating EBI metric and its geometric structure, including detailed analyses of the event horizon, static limit, and ergoregion, focusing on the influence of the Born--Infeld parameter. In Section \ref{S3}, we derive equations of motion for neutral particles and study their kinematics in the modified geometry, emphasising angular velocity and the radial effective potential under Born--Infeld corrections. Section~\ref{S4} examines the formation of negative energy orbits and the energy extraction via the Penrose process, highlighting how Born--Infeld nonlinearity affects efficiency and constraints, with implications from the Wald inequality.
 Section \ref{S7} analyses the irreducible mass dependence on the Born--Infeld parameter, presenting numerical trends. Finally, Section \ref{S8} summarises the main findings.

\section{Rotating Einstein--Born--Infeld Black Hole Metric} \label{S2}

The rotating EBI black hole is a solution to Einstein's field equations coupled with nonlinear Born--Infeld electrodynamics, generalizing the Kerr--Newman black hole by incorporating a Born--Infeld parameter \(\beta\). This parameter regulates the electromagnetic field strength and eliminates the divergence of self-energy near the singularity. Rotating regular black hole solutions such as these can be generated using physical and symmetry-based methods that avoid the Newman--Janis complexification procedure, as discussed in recent works \cite{azreg2014static, azreg2014regular}.

In Boyer--Lindquist coordinates \((t, r, \theta, \phi)\), the line element \cite{atamurotov2016horizon} is given by

\begin{equation}
\begin{split}
ds^2 = {} & \frac{\Delta_{\text{EBI}}(r) - a^2 \sin^2 \theta}{\rho^2} dt^2 - \frac{\rho^2}{\Delta_{\text{EBI}}(r)} dr^2 \\
          & + 2 a \sin^2 \theta \left(1 - \frac{\Delta_{\text{EBI}}(r) - a^2 \sin^2 \theta}{\rho^2}\right) dt\, d\phi - \rho^2 d\theta^2 \\
          & - \sin^2 \theta \left[\rho^2 + a^2 \sin^2 \theta \left(2 - \frac{\Delta_{\text{EBI}}(r) - a^2 \sin^2 \theta}{\rho^2}\right)\right] d\phi^2,
\end{split}
\label{Metric}
\end{equation}

where the metric functions are

\begin{equation}
\Delta_{\text{EBI}}(r) = r^2 + a^2 - 2Mr + Q_{\text{eff}}^2(r), \quad \rho^2 = r^2 + a^2 \cos^2 \theta.
\label{Delleq}
\end{equation}

Here, \(M\) is the black hole mass, \(a\) is the spin (angular momentum per unit mass) parameter, \(Q\) is the electric charge, \(\beta\) the Born--Infeld parameter characterizing nonlinear electrodynamics, and \(G\) is Newton’s gravitational constant.

The effective radial-dependent squared charge \(Q_{\text{eff}}^2(r)\), arising from the nonlinear Born--Infeld coupling, is given by

\begin{equation}
Q_{\text{eff}}^2(r) = \frac{2 \beta^2 r^4}{3} \left(1 - \sqrt{1 + \frac{Q^2}{\beta^2 r^4}}\right) + \frac{4 Q^2}{3} \, {}_2F_1\left(\frac{1}{4}, \frac{1}{2}, \frac{5}{4}, -\frac{Q^2}{\beta^2 r^4}\right),
\label{Qsquare}
\end{equation}

where \({}_2F_1\) is the Gaussian hypergeometric function. A detailed discussion of ${}_2F_1$ can be found in standard references on special functions \cite{abramowitz1964handbook}.

In the Maxwell limit \(\beta \to \infty\), the effective charge reduces to the constant \(Q^2\), recovering the Kerr--Newman metric. For neutral black holes (\(Q=0\)), this reduces to the Kerr solution. The Born--Infeld parameter \(\beta\) modifies classical results by smoothing electromagnetic singularities near the horizon, affecting the horizon structure, ergoregions, and observables such as black hole shadows.

\subsection{Geometric Structure of the EBI Spacetime}

The horizons of rotating EBI black holes are determined by the roots of the metric function \(\Delta_{\text{EBI}}(r)=0\)
, where \(\Delta_{\text{EBI}}(r)\)
 incorporates a nonlinear dependence on the Born--Infeld parameter $\beta$ through an effective charge function $Q^2_{\text{eff}}(r)$. Owing to this nonlinear electrodynamics, the condition \(\Delta_{\text{EBI}}(r)=0\)
 leads to a transcendental equation that generally cannot be solved analytically and must be handled numerically.

Typically, the spacetime admits two distinct horizons: the outer event horizon and the inner Cauchy horizon. These horizons approach each other as extremality is approached and eventually coincide at the extremal limit. In the classical limit $\beta \to \infty$, the nonlinear effects vanish, and the solution smoothly reduces to that of the Kerr--Newman black hole.

\subsubsection{Parameter Space and Horizon Formation}

\begin{figure}[htbp]
    \centering
    \includegraphics[width=0.45\textwidth]{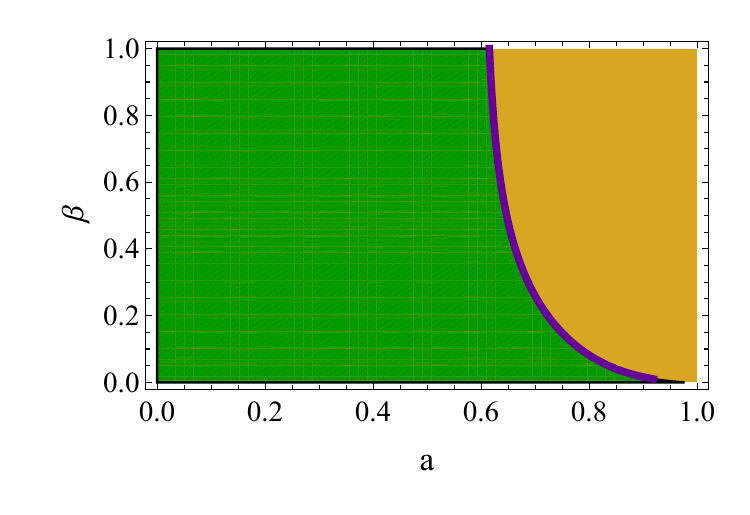}\hfill
    \includegraphics[width=0.45\textwidth]{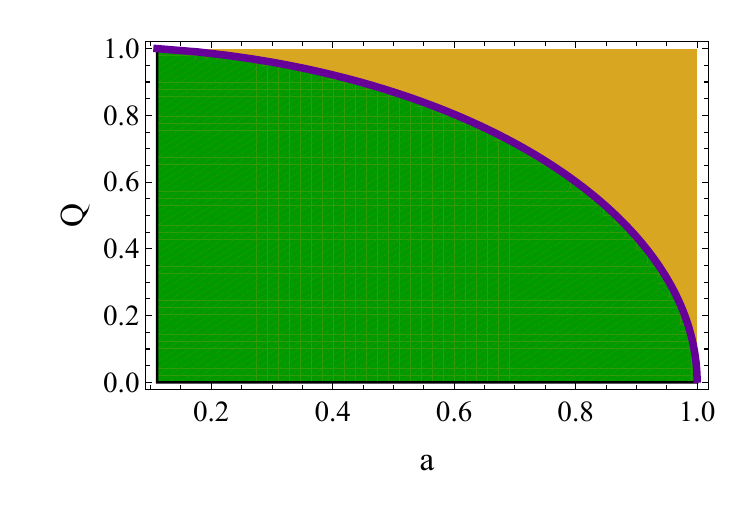}

    \vspace{0.2cm}

    \includegraphics[width=0.25\textwidth]{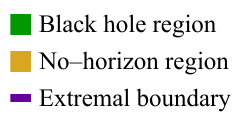}

    \caption{Parameter space of rotating EBI black holes showing spin $a$ versus charge $Q$ (left) and spin $a$ versus Born--Infeld parameter $\beta$ (right).}
    \label{fig:1HS}
\end{figure}

The parameter-space plots in Fig. \textbf{\ref{fig:1HS}} illustrate how horizon formation in rotating EBI black holes depends on the rotation parameter $a$, charge $Q$, and nonlinearity parameter $\beta$. The green region in each diagram marks combinations of parameters for which a physical event horizon exists, whereas the yellow region corresponds to horizonless solutions. In the left panel, increasing either rotation $a$ or charge $Q$ drives the solution closer to the extremal boundary (depicted in purple), beyond which the horizon disappears. The right panel shows that at sufficiently large $\beta$, the location of the extremal boundary becomes nearly insensitive to further increases, and the horizon structure stabilizes. These results highlight how spin, charge, and nonlinear electromagnetic corrections collectively constrain the existence and stability of black hole horizons in the EBI model.

\subsubsection{Behavior of Event and Cauchy Horizons}

\begin{figure}[htbp]
    \centering

    \makebox[\textwidth][c]{
        \includegraphics[width=0.55\textwidth]{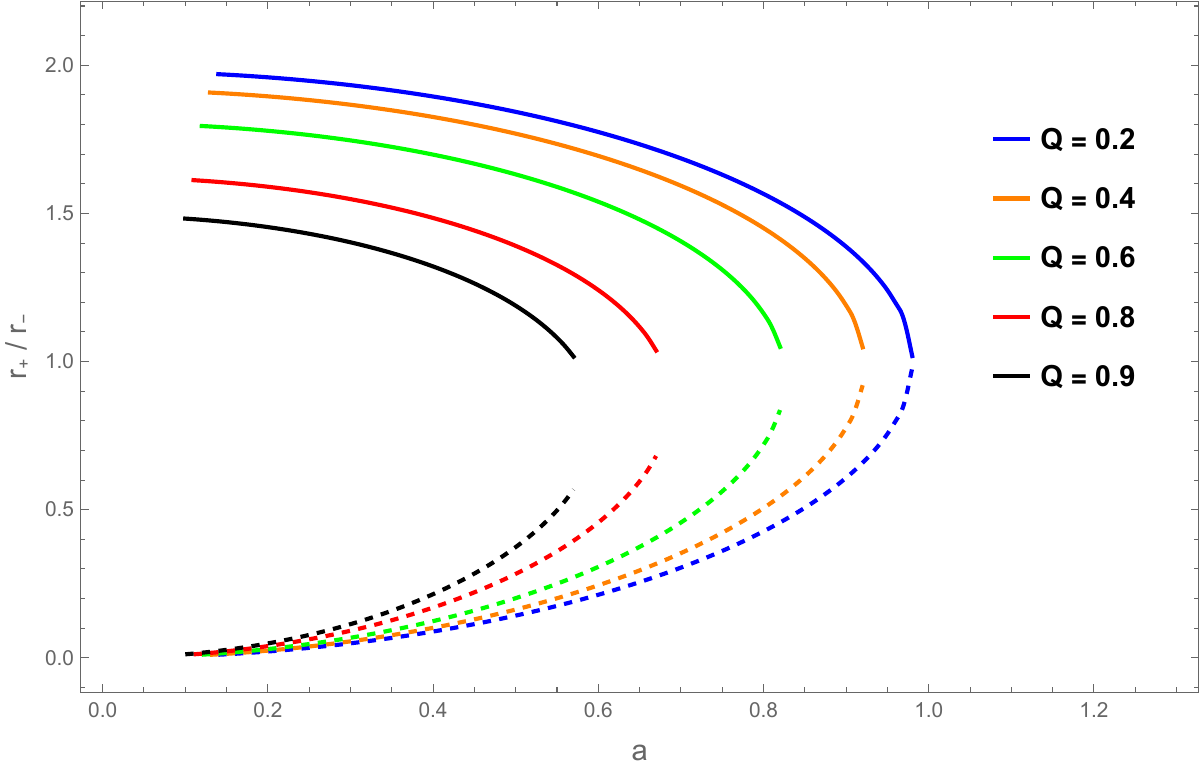}
        \hspace{2cm}
        \includegraphics[width=0.55\textwidth]{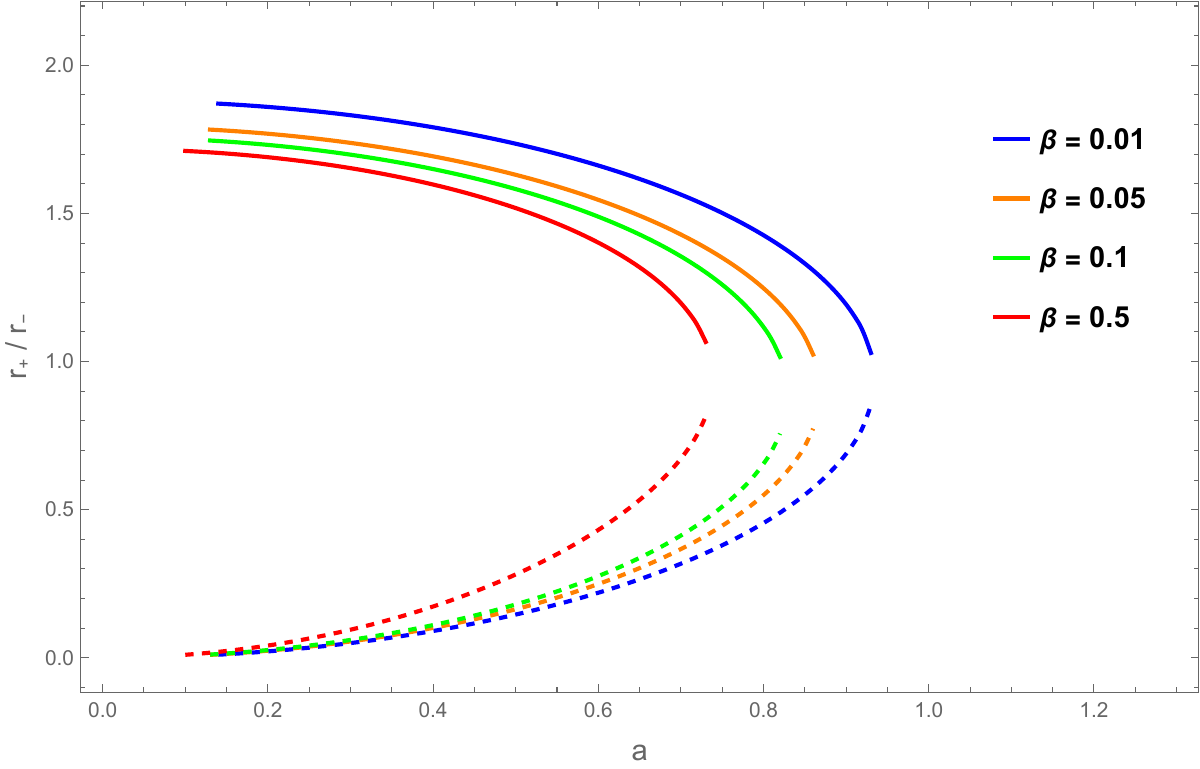}
    }

    \caption{Event (solid) and Cauchy (dotted) horizon radii versus spin $a$ for varying $Q$ (left) and $\beta$ (right).}
    \label{fig:2EChorizon}
\end{figure}

In Fig. \ref{fig:2EChorizon} the left panel shows the behavior of the event horizon (solid curves) and Cauchy horizon (dotted curves) radii as functions of the rotation parameter $a$ for various fixed values of the electric charge $Q$, with the Born--Infeld parameter $\beta$ held constant. Increasing $Q$ reduces the maximum spin at which both horizons exist, indicating that stronger charge effects drive the black hole closer to extremality. The event horizon radius decreases with increasing spin, while the Cauchy horizon approaches the event horizon radius as the extremal limit is approached.

The right panel illustrates how varying $\beta$, with fixed charge $Q$, affects the event and Cauchy horizons. Smaller $\beta$ values, corresponding to stronger nonlinear electromagnetic corrections, shrink the radii of both horizons and reduce the maximum allowed spin before horizon disappearance. As $\beta$ increases, the system approaches the linear Maxwell limit, maintaining larger horizon radii up to higher spins. These plots demonstrate the combined influence of charge and nonlinear electrodynamics on the horizon structure of rotating EBI black holes.

\subsubsection{Metric Function \texorpdfstring{$\Delta(r)$}{?(r)} and Horizon Roots}

\begin{figure}[htbp]
    \centering

    \begin{minipage}[b]{0.45\textwidth}
        \centering
      (a)  \includegraphics[width=\textwidth]{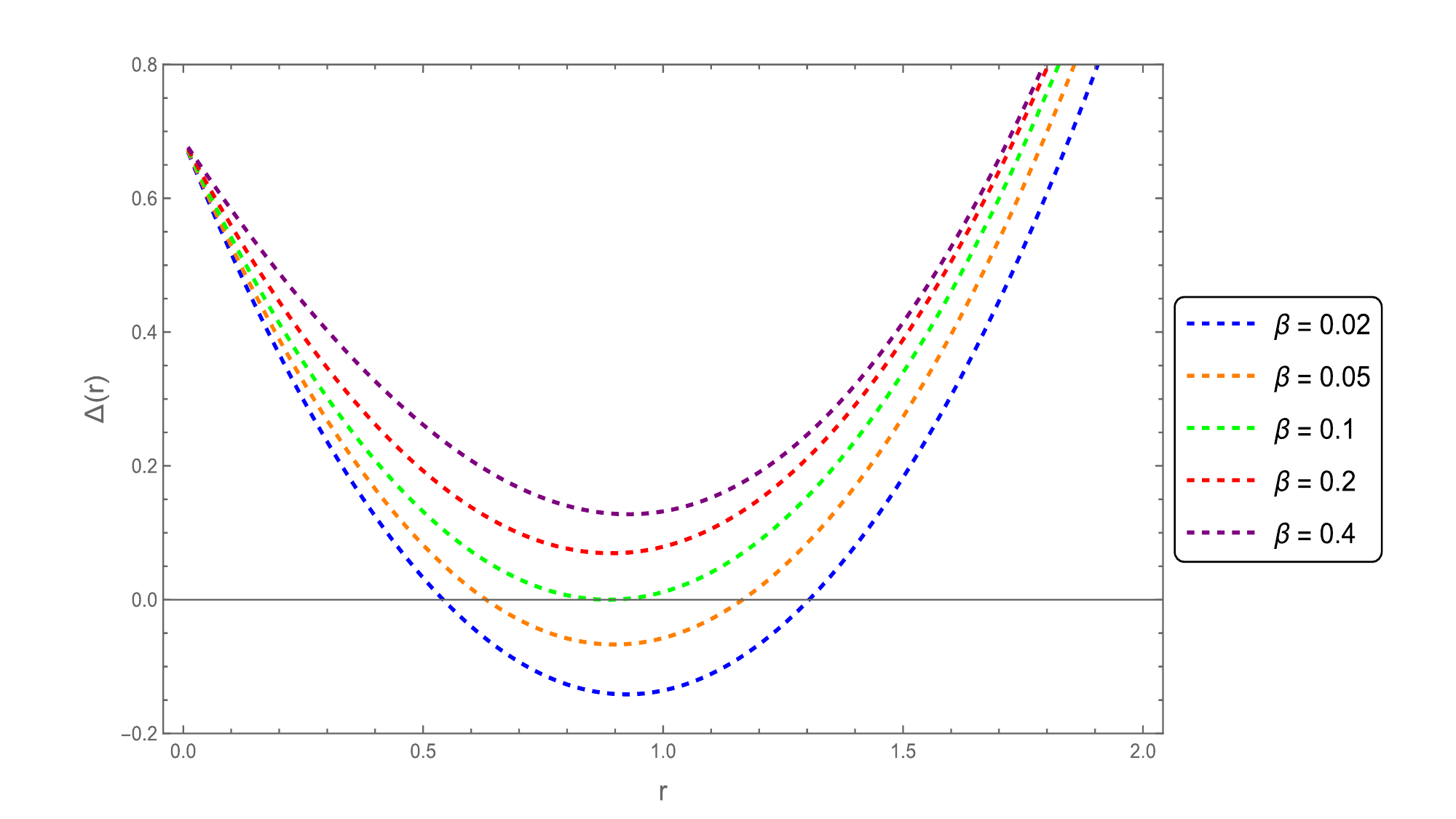}
    \end{minipage}
    \hspace{0.05\textwidth}
    \begin{minipage}[b]{0.45\textwidth}
        \centering
       (b) \includegraphics[width=\textwidth]{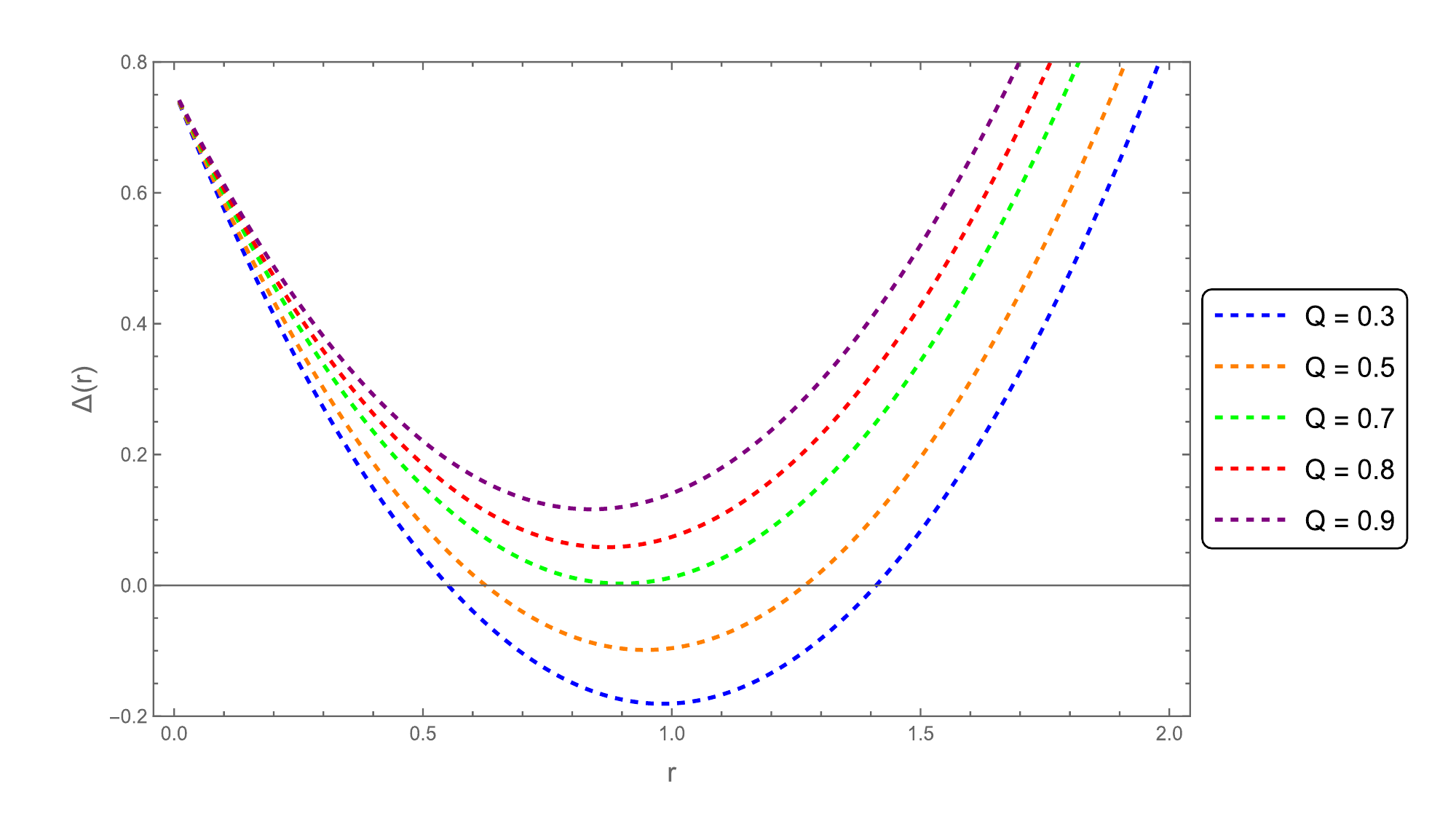}
    \end{minipage}

    \vspace{0.8cm} 

    \begin{minipage}[b]{0.45\textwidth}
        \centering
      (c)  \includegraphics[width=\textwidth]{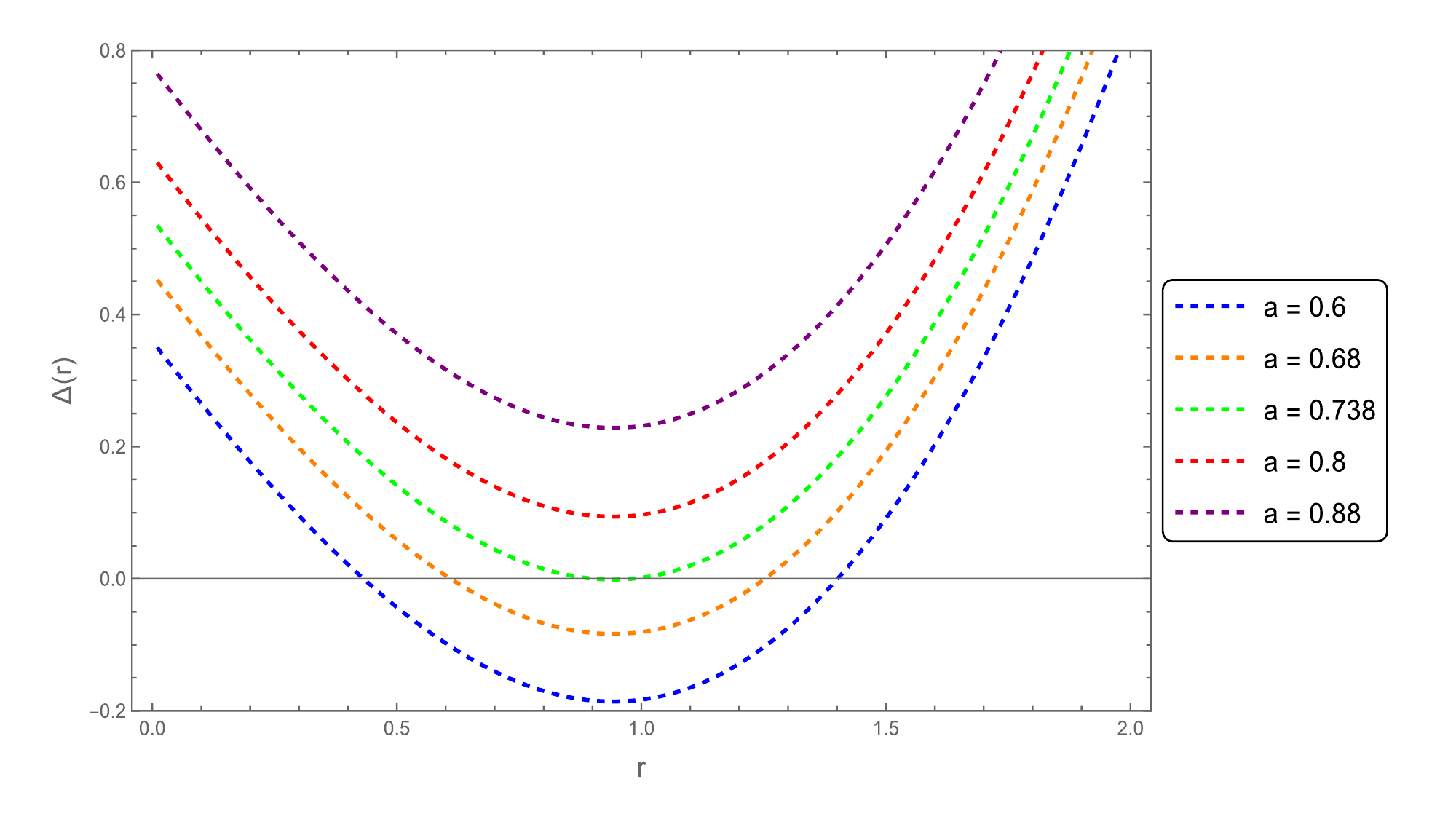}
    \end{minipage}

    \caption{Metric function \(\Delta(r)\) versus radial coordinate \(r\) for rotating EBI black holes.}
    \label{fig:3Dell(r)}
\end{figure}

The metric function \(\Delta(r)\) in Fig. \ref{fig:3Dell(r)} governs the horizon structure of rotating EBI black holes. Plot (a) and (b) show that increasing the Born--Infeld parameter \(\beta\) or electric charge \(Q\) raises the minimum of \(\Delta(r)\), reducing the radial range where \(\Delta(r)<0\) and thus narrowing the horizon-forming region. Smaller \(\beta\) and larger \(Q\) expand horizon existence, while higher values push the system toward extremality. Plot (c) demonstrates that increasing spin \(a\) raises and shifts the minimum of \(\Delta(r)\), limiting the parameter space for horizons and lowering the maximal spin sustaining them. These results highlight the combined effects of nonlinearity, charge, and rotation in shaping horizon geometry via the modulation of \(\Delta(r)\), emphasizing the crucial role of nonlinear electrodynamics in the stability and existence of rotating charged black hole horizons.

\subsubsection{Ergoregion Structure}

\begin{figure}[H]
    \centering
  (a) \includegraphics[width=1\textwidth]{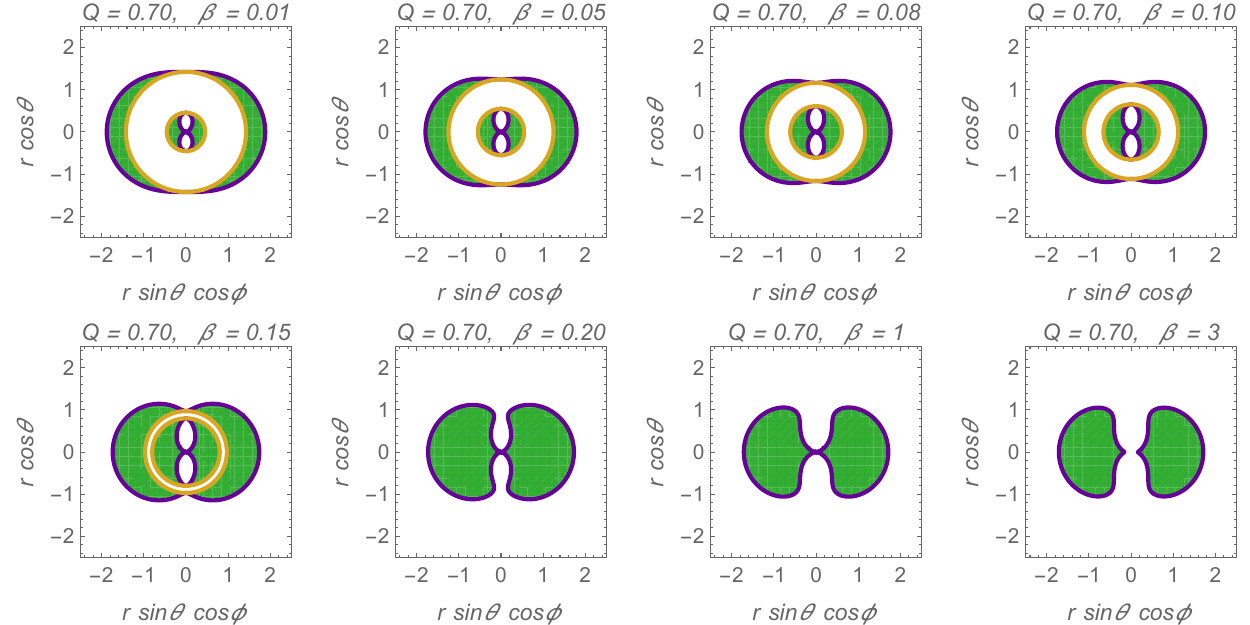}
    \label{fig:fig1}
    \vspace{1.5em}
  (b)  \includegraphics[width=1\textwidth]{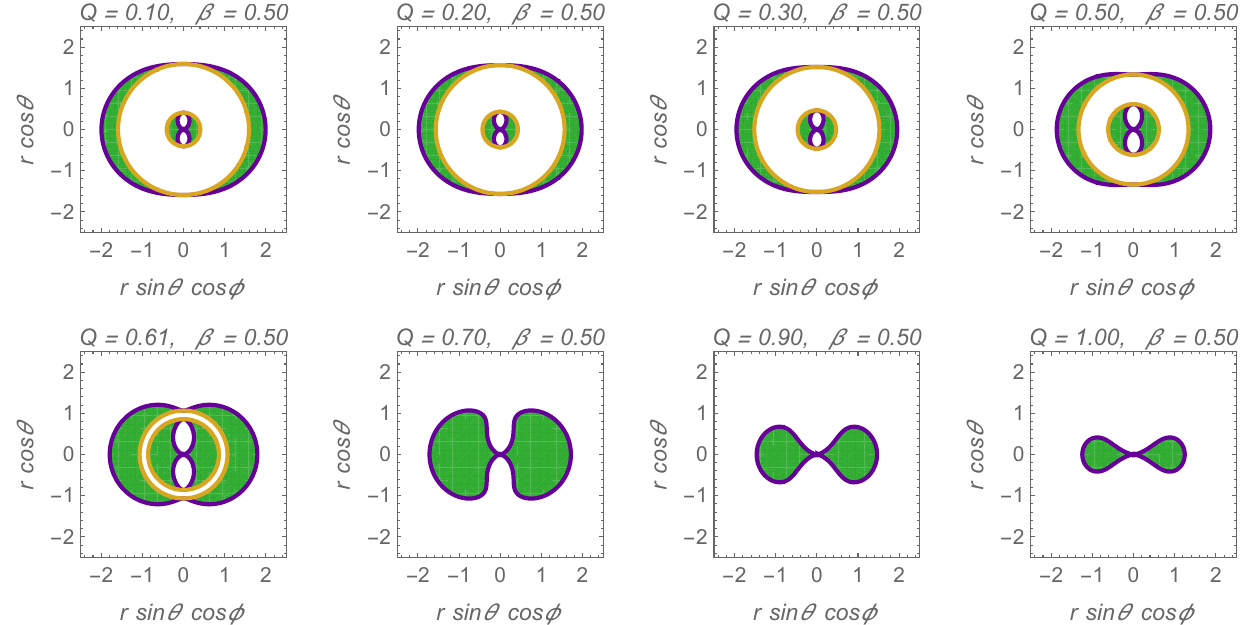}

\vspace{2em}

    \includegraphics[width=0.6\textwidth]{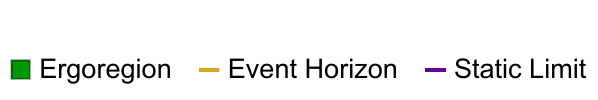}
     \caption{Cross-sectional views of the ergoregion, event horizon, and static limit for various  and $\beta$ and $Q$.}
      \label{fig:4ergoregion}
\end{figure}

Increasing the Born--Infeld parameter \(\beta\) at fixed charge and spin leads to a systematic contraction of both the event horizon radius \(r_+\) and the static limit radius \(r_{es}^+\) (see Table~\ref{T:ErgoregionBeta}). Despite this inward shift, the ergoregion width \(\delta = r_{es}^+ - r_+\) increases, indicating an expanded zone dominated by frame-dragging effects. This growth is evident in the cross-sectional plots of Fig.~\ref{fig:4ergoregion}(a), where larger \(\beta\) values produce a noticeably bigger ergoregion. At high \(\beta\), the ergoregion undergoes topological changes, splitting into multiple disconnected parts near the equatorial plane, revealing strong nonlinear electromagnetic influences on the spacetime geometry.

Similarly, with fixed spin, increasing the electric charge \(Q\) causes decreases in both the event horizon \(r_+\) and static limit \(r_{es}^+\) radii, as detailed in Table~\ref{T:ErgoregionQ}. The ergoregion width \(\delta\) correspondingly grows, reflecting an enlarged frame-dragging region. Figure~\ref{fig:4ergoregion}(b) illustrates these effects by showing progressively more complex ergoregion shapes with increasing \(Q\). For sufficiently high charge, the ergoregion splits into multiple disconnected lobes, highlighting the intricate interplay between charge and nonlinear electromagnetic effects shaping the spacetime structure.

\begin{table}[h!]
\centering
\caption{The ergoregion size ($\delta = r_{\text{es}}^+ - r_+$), event horizon ($r_+$), and static limit ($r_{\text{es}}^+$) for different spin values $a$ and Born--Infeld parameter $\beta$.}
\resizebox{\textwidth}{!}{%
\begin{tabular}{|c|ccc|ccc|ccc|ccc|}
\hline
\multirow{2}{*}{$\beta$} &
\multicolumn{3}{c|}{$a = 0.5$} &
\multicolumn{3}{c|}{$a = 0.6$} &
\multicolumn{3}{c|}{$a = 0.7$} &
\multicolumn{3}{c|}{$a = 0.8$} \\
\cline{2-13}
& $r_+$ & $r_{\text{es}}^+$ & $\delta$
& $r_+$ & $r_{\text{es}}^+$ & $\delta$
& $r_+$ & $r_{\text{es}}^+$ & $\delta$
& $r_+$ & $r_{\text{es}}^+$ & $\delta$ \\
\hline
0.01 & 1.7905 & 1.8637 & 0.0732 & 1.7202 & 1.8323 & 0.1121 & 1.6275 & 1.7936 & 0.1660 & 1.5010 & 1.7465 & 0.2455 \\
0.05 & 1.7399 & 1.8176 & 0.0778 & 1.6644 & 1.7843 & 0.1199 & 1.5634 & 1.7431 & 0.1797 & 1.4266 & 1.6928 & 0.2722 \\
0.08 & 1.7273 & 1.8068 & 0.0795 & 1.6499 & 1.7728 & 0.1229 & 1.5454 & 1.7306 & 0.1853 & 1.3952 & 1.6790 & 0.2838 \\
0.10 & 1.7224 & 1.8027 & 0.0803 & 1.6440 & 1.7684 & 0.1243 & 1.5379 & 1.7258 & 0.1879 & 1.3838 & 1.6736 & 0.2897 \\
0.15 & 1.7157 & 1.7972 & 0.0815 & 1.6358 & 1.7624 & 0.1266 & 1.5267 & 1.7191 & 0.1924 & 1.3657 & 1.6659 & 0.3002 \\
0.20 & 1.7125 & 1.7947 & 0.0822 & 1.6317 & 1.7596 & 0.1279 & 1.5209 & 1.7160 & 0.1950 & 1.3553 & 1.6622 & 0.3069 \\
1.00 & 1.7074 & 1.7908 & 0.0834 & 1.6249 & 1.7552 & 0.1303 & 1.5105 & 1.7109 & 0.2004 & 1.3331 & 1.6561 & 0.3229 \\
3.00 & 1.7071 & 1.7906 & 0.0835 & 1.6245 & 1.7550 & 0.1305 & 1.5100 & 1.7107 & 0.2007 & 1.3318 & 1.6558 & 0.3240 \\
\hline
\end{tabular}%
}
\label{T:ErgoregionBeta}
\end{table}

\begin{table}[h!]
\centering
\caption{The ergoregion size ($\delta = r_{\text{es}}^+ - r_+$), event horizon ($r_+$), and static limit ($r_{\text{es}}^+$) for different spin values $a$ and charge $Q$.}
\resizebox{\textwidth}{!}{%
\begin{tabular}{|c|ccc|ccc|ccc|ccc|}
\hline
\multirow{2}{*}{$Q$} &
\multicolumn{3}{c|}{$a = 0.5$} &
\multicolumn{3}{c|}{$a = 0.6$} &
\multicolumn{3}{c|}{$a = 0.7$} &
\multicolumn{3}{c|}{$a = 0.8$} \\
\cline{2-13}
& $r_+$ & $r_{\text{es}}^+$ & $\delta$
& $r_+$ & $r_{\text{es}}^+$ & $\delta$
& $r_+$ & $r_{\text{es}}^+$ & $\delta$
& $r_+$ & $r_{\text{es}}^+$ & $\delta$ \\
\hline
0.1 & 1.8603 & 1.9301 & 0.0698 & 1.7938 & 1.9000 & 0.1063 & 1.7072 & 1.8632 & 0.1560 & 1.5917 & 1.8186 & 0.2269 \\
0.2 & 1.8430 & 1.9141 & 0.0711 & 1.7751 & 1.8835 & 0.1084 & 1.6862 & 1.8460 & 0.1597 & 1.5667 & 1.8004 & 0.2337 \\
0.3 & 1.8143 & 1.8875 & 0.0732 & 1.7440 & 1.8561 & 0.1121 & 1.6513 & 1.8174 & 0.1660 & 1.5248 & 1.7703 & 0.2455 \\
0.4 & 1.7742 & 1.8504 & 0.0762 & 1.7005 & 1.8177 & 0.1173 & 1.6022 & 1.7774 & 0.1752 & 1.4648 & 1.7281 & 0.2633 \\
0.5 & 1.7224 & 1.8027 & 0.0803 & 1.6441 & 1.7684 & 0.1243 & 1.5379 & 1.7258 & 0.1879 & 1.3838 & 1.6736 & 0.2897 \\
0.6 & 1.6536 & 1.7441 & 0.0855 & 1.5740 & 1.7077 & 0.1337 & 1.4566 & 1.6623 & 0.2056 & 1.2750 & 1.6060 & 0.3310 \\
0.7 & 1.5820 & 1.6743 & 0.0923 & 1.4889 & 1.6351 & 0.1463 & 1.3549 & 1.5859 & 0.2310 & 1.1164 & 1.5244 & 0.4079 \\
0.8 & 1.4913 & 1.5925 & 0.1012 & 1.3862 & 1.5499 & 0.1636 & 1.2251 & 1.4957 & 0.2706 & -- & -- & -- \\
0.9 & 1.3846 & 1.4980 & 0.1133 & 1.2616 & 1.4506 & 0.1890 & 1.0435 & 1.3896 & 0.3461 & -- & -- & -- \\
1.0 & 1.2587 & 1.3893 & 0.1306 & 1.1046 & 1.3356 & 0.2309 & -- & -- & -- & -- & -- & -- \\
\hline
\end{tabular}%
}
\label{T:ErgoregionQ}
\end{table}

\section{Particle Motion in the Spacetime of a Rotating Einstein--Born--Infeld Black Hole} \label{S3}

To investigate the dynamics of neutral particles near a rotating EBI black hole, we consider their trajectories confined to the equatorial plane (\(\theta = \pi/2\)), a region of astrophysical interest due to its symmetry and observational relevance. The test particle paths, or timelike geodesics, are fully determined by the EBI spacetime geometry.

Using the Lagrangian formalism, the motion of a unit rest mass particle is described by
\begin{equation}
\mathcal{L} = \frac{1}{2} g_{\mu\nu} \dot{x}^\mu \dot{x}^\nu,
\end{equation}
where the overdot indicates differentiation with respect to the particle’s proper time $\tau$.

Because the spacetime is stationary and axisymmetric, the coordinates \( t \) and \( \phi \) are cyclic, leading to conserved quantities: the particle energy \( E \) and angular momentum \( L \), expressed as

\begin{align}
-p_t &= g_{tt}\dot{t} + g_{t\phi}\dot{\phi} = E, \label{pt_eq} \\
p_\phi &= g_{t\phi}\dot{t} + g_{\phi\phi}\dot{\phi} = L, \label{pphi_eq}
\end{align}

Solving the above system of Eqns. (\ref{pt_eq}) and (\ref{pphi_eq}) for \(\dot{t}\) and \(\dot{\phi}\), we find:
\begin{align}
\dot{t} &= \frac{1}{r^2 \Delta_{\text{EBI}}} \left[ E a^2 ( Q^2(r) - r(2M + r)) + a L ( Q^2(r) - 2Mr) - E r^4 \right], \label{a4} \\
\dot{\phi} &= \frac{1}{r^2 \Delta_{\text{EBI}}} \left[ a E ( Q^2(r) - 2Mr) + L ( Q^2(r) + r(r - 2M)) \right] \label{a5}.
\end{align}

In the rotating EBI black hole case, these expressions include modifications from nonlinear electrodynamics, reflected in the altered radial function \(\Delta_{\text{EBI}}(r)\)
, distinguishing the spacetime from the classical Kerr and Kerr--Newman solutions.

To further describe the radial behavior, the Hamiltonian formalism is employed, offering a phase-space perspective on geodesic motion. The Hamiltonian for a unit-mass particle reads
\begin{equation}
\mathcal{H} = \frac{1}{2} g^{\mu\nu} p_\mu p_\nu,
\end{equation}
with $g^{\mu\nu}$ denoting the contravariant metric tensor and $p_\mu$ the covariant momenta. Timelike geodesics satisfy the normalization condition
\begin{equation}
2\mathcal{H} = \epsilon = -1. \label{normalization}
\end{equation}

Using the conserved quantities and normalization condition (\ref{normalization}), the radial equation of motion takes the form
\begin{equation}
2\mathcal{H} = E \dot{t} + L \dot{\phi} + \frac{r^2}{\Delta_{\text{EBI}}} \dot{r}^2 = \epsilon,
\label{finalnorm}
\end{equation}

Solving Eq. (\ref{finalnorm}) for \(\dot{r}^2\), we obtain the radial equation
\begin{equation}
\dot{r}^2 = E^2 + \frac{(2Mr -  Q^2(r))(aE + L)^2}{r^4} + \frac{1}{r^2} \left( a^2 E^2 - L^2 \right) + \epsilon  \frac{\Delta_{\text{EBI}}}{r^2}. \label{radial:eq}
\end{equation}

Rearranging, we obtain an explicit expression for \( \dot{r}^2 \) that shows how the radial velocity depends on the particle’s energy, angular momentum, and the Born--Infeld-corrected metric functions. This radial equation captures the impact of the EBI parameter on particle trajectories and provides a foundation for analyzing orbital stability, locating circular orbits, and investigating energy extraction processes.

Another key aspect governing particle dynamics in the vicinity of a rotating EBI black hole is the angular velocity profile. This quantity characterizes the rate at which a particle’s azimuthal coordinate $\phi$ changes with respect to the coordinate time $t$, as measured by a distant observer. Due to the black hole's rotation, frame dragging compels particles to partially co-rotate with the black hole, an effect that intensifies inside the ergoregion \cite{bardeen1972rotating}.

The angular velocity is defined as
\begin{equation}
\Omega^{\text{EBI}} = \frac{d\phi}{dt}, \label{eq:ebi-omega-def}
\end{equation}
and its physically allowed values are constrained by the geometry of spacetime. By imposing the condition for timelike trajectories, $ds^2 \geq 0$, one obtains the permissible range:
\begin{equation}
\Omega^{\text{EBI}}_{\pm} = \frac{-g_{t\phi} \pm \sqrt{g_{t\phi}^2 - g_{tt} g_{\phi\phi}}}{g_{\phi\phi}}, \label{eq:ebi-omega-bounds}
\end{equation}
where the signs ``$+$" and ``$-$'' correspond to the maximum and minimum allowed angular velocities, respectively.

At the static limit surface where $g_{tt} = 0$, the upper bound $\Omega^{\text{EBI}}_{+}$ vanishes. However, within the ergoregion, both $\Omega^{\text{EBI}}_{+}$ and $\Omega^{\text{EBI}}_{-}$ are positive, indicating that all particles are forced to co-rotate with the black hole’s spin.

Substituting the rotating EBI black hole metric components into Eq.~\eqref{eq:ebi-omega-bounds} leads to
\begin{equation}
\Omega^{\text{EBI}}_{\pm} = \omega^{\text{EBI}} \pm \frac{r^2 \sqrt{\Delta_{\text{EBI}}(r)}}{(a^2 + r^2)^2 - a^2 \Delta_{\text{EBI}}(r)},
\end{equation}
with the frame-dragging angular velocity given by
\begin{equation}
\omega^{\text{EBI}} = \frac{a \left(2 M r - Q_{\text{eff}}^2(r) \right)}{(a^2 + r^2)^2 - a^2 \Delta_{\text{EBI}}(r)},
\end{equation}
where the function $\Delta_{\text{EBI}}(r)$ characterizes the location of horizons in the EBI spacetime.

In the near-horizon limit $r \to r_+$, where $r_+$ denotes the event horizon radius, the angular velocity bounds converge smoothly
\begin{equation}
\lim_{r \to r_+} \Omega^{\text{EBI}}_{\pm} = \Omega_{\text{H}}^{\text{EBI}} = \frac{a \left(2 M r_+ - Q_{\text{eff}}^2(r_+) \right)}{r_+^4 + a^2 r_+^2 - a^2 \left(2 M r_+ - Q_{\text{eff}}^2(r_+)\right)},
\end{equation}
confirming that all particles are inevitably dragged to co-rotate with the black hole at the horizon.

Restricting to the equatorial plane $\theta = \pi/2$, the angular velocity in terms of the conserved specific energy $E$ and angular momentum $L$ reads
\begin{equation}
\Omega^{\text{EBI}} = \frac{a \left(2 M r^3 - Q_{\text{eff}}^2(r) r^2 \right) E - \left(r^4 - 2 M r^3 + Q_{\text{eff}}^2(r) r^2\right) L}{\left[r^4 (r^2 + a^2) + a^2 \left(2 M r^3 - Q_{\text{eff}}^2(r) r^2 \right)\right] E + a L \left(2 M r^3 - Q_{\text{eff}}^2(r) r^2\right)}. \label{eq:ebi-omega-full}
\end{equation}

The conserved energy and angular momentum for circular orbits satisfy
\begin{align}
E &= \frac{r^4 - 2 M r^3 + Q_{\text{eff}}^2(r) r^2 \pm a r^2 \sqrt{M r - Q_{\text{eff}}^2(r)}}
{r^3 \sqrt{r^4 - 3 M r^3 +2 Q_{\text{eff}}^2(r) r^2 \pm 2 a r^2 \sqrt{M r - {Q_{\text{eff}}^2(r)}}}}, \label{eq:E-circular}\\[8pt]
L &= \pm \frac{r^4 + a^2 r^2 \mp 2 a r^2 \sqrt{M r - {Q_{\text{eff}}^2(r)}} \mp a Q_{\text{eff}}^2(r) r^2}
{r^2 \sqrt{r^4 - 3 M r^3 +2 Q_{\text{eff}}^2(r) r^2 \pm 2 a r^2 \sqrt{M r - Q_{\text{eff}}^2(r)}}}. \label{eq:L-circular}
\end{align}
Here, the upper and lower signs correspond to co-rotating and counter-rotating orbits, respectively.

Substituting Eqs.~\eqref{eq:E-circular} and \eqref{eq:L-circular} into Eq.~\eqref{eq:ebi-omega-full} simplifies the angular velocity to the compact form
\begin{equation}
\Omega^{\text{EBI}} = \frac{\mp r \sqrt{M r^3 - Q_{\text{eff}}^2(r) r^2}}{r^3 \mp a \sqrt{M r^3 - Q_{\text{eff}}^2(r) r^2}},
\end{equation}
recovering the expected dependence on black hole spin $a$, radial coordinate $r$, and the nonlinear charge parameter $Q_{\text{eff}}(r)$. Setting $a = 0$ retrieves the results for a static (non-rotating) EBI black hole, where frame dragging vanishes and angular velocity reduces accordingly.

\begin{figure}[htbp]
    \centering

    \begin{minipage}{0.45\textwidth}
        \centering
        \includegraphics[width=\linewidth]{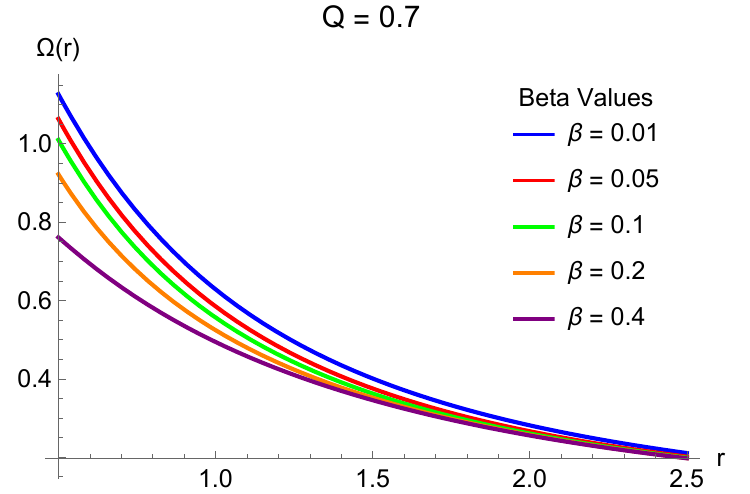}
    \end{minipage}%
    \hfill
    \begin{minipage}{0.45\textwidth}
        \centering
        \includegraphics[width=\linewidth]{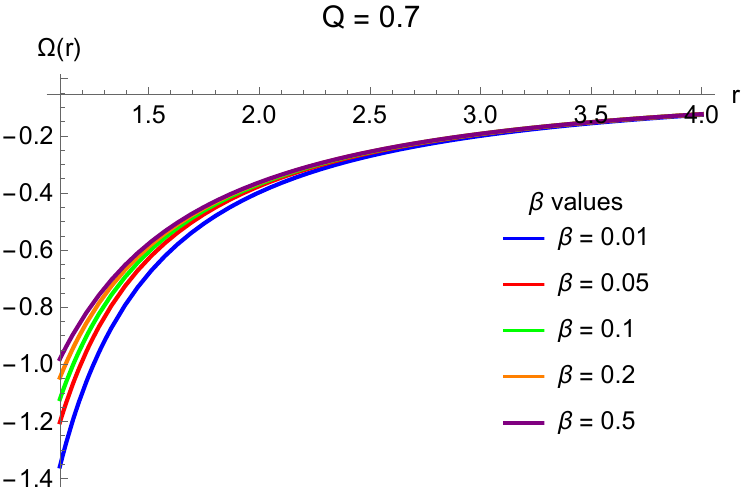}
    \end{minipage}

    \vspace{1em} 

    \begin{minipage}{0.45\textwidth}
        \centering
        \includegraphics[width=\linewidth]{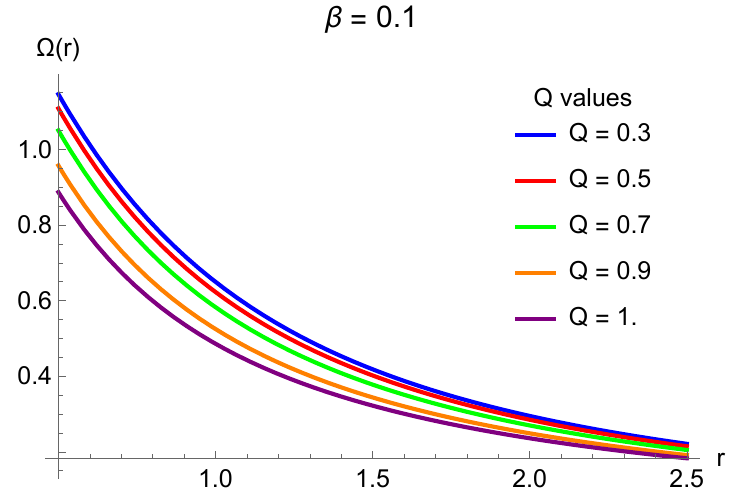}
    \end{minipage}%
    \hfill
    \begin{minipage}{0.45\textwidth}
        \centering
        \includegraphics[width=\linewidth]{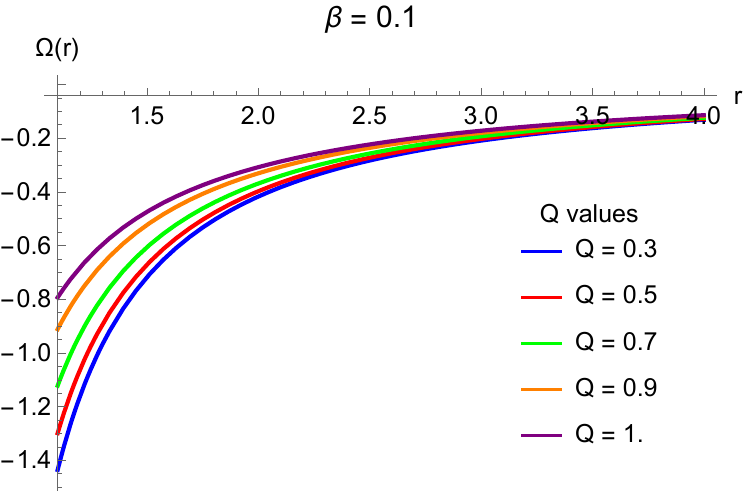}
    \end{minipage}

    \caption{Dependence of angular velocity $\Omega$ on $Q$  and $\beta$ versus radial distance $r$.}
    \label{fig:5Angvelocity}
\end{figure}

The angular velocity $\Omega(r)$ in Fig. \ref{fig:5Angvelocity} of a rotating EBI black hole decreases with radial distance $r$, and both increasing the Born--Infeld parameter $\beta$ and the charge $Q$ lower $\Omega(r)$ near the black hole, with their effects diminishing at large distances. This indicates that higher nonlinearity and charge reduce frame-dragging effects close to the black hole, while far from it, the influence of these parameters becomes negligible.

\section{Collisional Energy Extraction in the Quantum-Corrected Ergoregion} \label{S4}

We follow the notation of Sec.~\textbf{\ref{S2} }(Eqs.~\eqref{Metric}–\eqref{Qsquare}).
Nonlinear Born–Infeld electrodynamics modifies the Kerr geometry through the radius–dependent charge \(Q_{\rm eff}^2(r)\), which enters both \(\Delta_{EBI}(r)\) and \(g_{tt}\).
Consequently, the location and size of the ergoregion (\(g_{tt}>0\)) depend on \((a,Q,\beta)\), shifting the kinematic bounds on the conserved quantities \(E=-u_t\) and \(L=u_\phi\).
Inside the ergoregion, counter–rotating fragments can realize negative energy states \((E<0)\), and in collisional (or fission) setups this enables \(E_{\rm out}>E_{\rm in}\) and thus energy extraction.
Below we recast the EBI radial potential into explicit bounds \(E(r,L)\) and \(L(r,E)\) and evaluate the resulting collisional efficiency.

We consider neutral test particles in the equatorial plane of a rotating EBI black hole. Here we are using
\begin{align}
\Gamma_{\text{EBI}}(r) &\equiv 2Mr - Q_{\text{eff}}^2(r),\qquad
g_{tt} = -\!\left(1-\frac{\Gamma_{\text{EBI}}}{r^2}\right).
\end{align}
For completeness, the Born--Infeld effective charge function \(Q_{\text{eff}}^2\!\to Q^2\) as \(\beta\!\to\!\infty\).

\subsection{Negative Energy States}

Inside the ergoregion (\(g_{tt}>0 \iff \Gamma_{\text{EBI}}(r)>r^2\)), counter-rotating geodesics can carry energy \(E\equiv -u_t<0\) as seen at infinity, enabling the Penrose mechanism. Using the radial equation (\ref{radial:eq})
with \(\epsilon=-1,0,+1\) for timelike, null, and spacelike motion, one obtains a quadratic relation
\begin{equation}
A_{\text{EBI}}\,E^2-2a\,\Gamma_{\text{EBI}}\,L\,E+\big(a^2-\Delta_{\text{EBI}}\big)L^2+\epsilon\,r^2\Delta_{\text{EBI}}=0,
\qquad
\mathrm{where}\quad  A_{\text{EBI}}(r)= r^4+a^2\!\big(r^2+\Gamma_{\text{EBI}}\big).
\label{eq:quad-EBI}
\end{equation}

\begin{figure}[htbp]
    \centering

    \begin{minipage}[b]{0.45\textwidth}
        \centering
      (a)  \includegraphics[width=\textwidth]{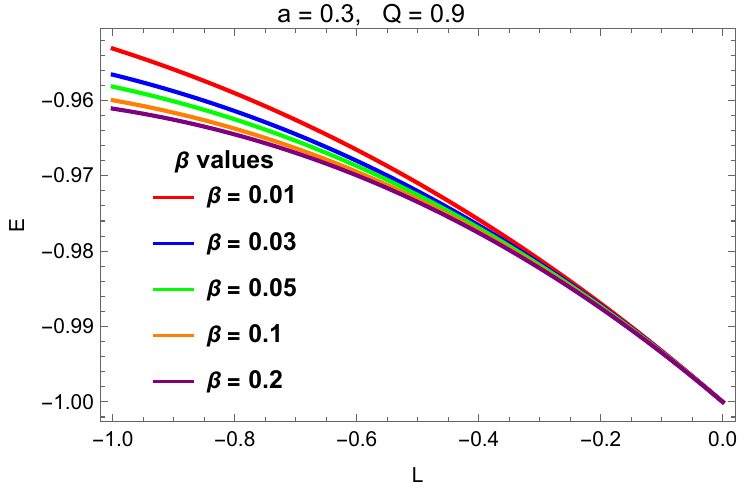}
    \end{minipage}
    \hspace{0.05\textwidth}
    \begin{minipage}[b]{0.45\textwidth}
        \centering
       (b) \includegraphics[width=\textwidth]{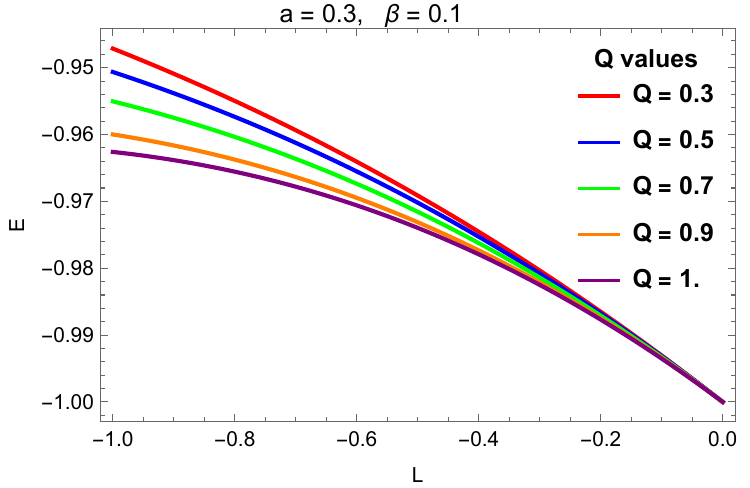}
    \end{minipage}

    \vspace{0.8cm} 

    \begin{minipage}[b]{0.45\textwidth}
        \centering
      (c)  \includegraphics[width=\textwidth]{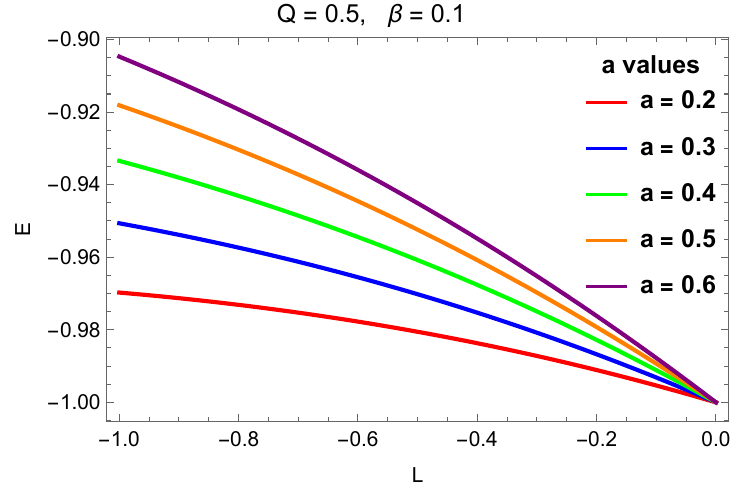}
    \end{minipage}

    \caption{The relationship between energy \(E\) and angular momentum \(L\) in negative energy states.}

    \label{fig:Negative energy}
\end{figure}

Solving \eqref{eq:quad-EBI} either for \(E\) or \(L\) gives
\begin{align}
E(r,L) &= \frac{a\,\Gamma_{\text{EBI}}\,L \;\pm\;
\sqrt{\,a^2\Gamma_{\text{EBI}}^2 L^2 - A_{\text{EBI}}\!\left[\big(a^2-\Delta_{\text{EBI}}\big)L^2+\epsilon\,r^2\Delta_{\text{EBI}}\right]}\,}{A_{\text{EBI}}}, \label{eq:EofL-EBI}\\
L(r,E) &= \frac{a\,\Gamma_{\text{EBI}}\,E \;\pm\;
\sqrt{\,a^2\Gamma_{\text{EBI}}^2 E^2 - \big(a^2-\Delta_{\text{EBI}}\big)\!\left(A_{\text{EBI}}E^2+\epsilon\,r^2\Delta_{\text{EBI}}\right)}\,}{\,a^2-\Delta_{\text{EBI}}\,}. \label{eq:LofE-EBI}
\end{align}
For timelike geodesics (\(\epsilon=-1\)), negative energy is permitted only where \(g_{tt}>0\) and is favored by counter-rotation \(aL<0\). A useful identity,
\begin{equation}
r^8\Delta_{\text{EBI}} - a^2\big(2Mr^3 - r^2 Q_{\text{eff}}^2\big)^2
= r^2\!\left[r^6 + a^2\!\big(r^4+2Mr^3 - r^2 Q_{\text{eff}}^2\big)\right]\!
\left[r^4 - 2Mr^3 + r^2 Q_{\text{eff}}^2\right],
\label{eq:identity-EBI}
\end{equation}
To determine when negative energy states (\(E < 0\)) are allowed, we set \(E = 1\) and analyze the conditions under which the resulting expression is negative. It follows clearly that negative energy necessitates \(L < 0\), and additional constraints can be derived by further simplifying the inequality.
\begin{equation}
a^2 L^2 \left( 2 M r^3 - \alpha M^2 \right)^2 >
r^2 \Delta \left[ r^4 L^2 - \epsilon \left( r^6 + a^2 \left( r^4 + 2 M r^3 - \alpha M^2 \right) \right) \right]. \label{Inequality}
\end{equation}
By using Eq. (\ref{eq:identity-EBI}), expression (\ref{Inequality}) takes the form
\begin{equation}
\frac{1}{r^4}\!\left[r^6 + a^2\!\big(r^4+2Mr^3 - r^2 Q_{\text{eff}}^2\big)\right]
\left[\big(r^4 - 2Mr^3 + r^2 Q_{\text{eff}}^2\big)L^2 - \epsilon\,r^4\Delta_{\text{EBI}}\right] < 0.
\label{eq:ineq-EBI}
\end{equation}
From \eqref{eq:ineq-EBI} one reads the practical criterion
\begin{equation}
E<0 \iff L<0\ \ \text{and}\ \
\frac{r^4 - 2Mr^3 + r^2 Q_{\text{eff}}^2}{r^4} \;<\; \frac{\epsilon\,\Delta_{\text{EBI}}}{L^2}.
\label{eq:negE-criterion-EBI}
\end{equation}

The graphs in Fig. \ref{fig:Negative energy} energy states $E$ versus angular momentum $L$. Increasing the Born--Infeld parameter $\beta$ (plot a) and charge $Q$ (plot b) raises the energy levels, reducing the depth of negative energy states. In contrast, increasing the spin $a$ (plot c) lowers the energy, enhancing the negative energy region. Thus, higher $\beta$ and $Q$ weaken, while greater spin strengthens, the negative energy states.

\subsection{Wald Inequality}

The Wald inequality sets bounds on fragment energies produced by a particle decaying inside a rotating black hole’s ergoregion, establishing kinematic conditions for Penrose energy extraction~\cite{wald1974energy,parthasarathy1986high}. Considering a parent particle with specific energy \(E = -\xi_\mu U^\mu,\) and a fragment with energy \(\epsilon = -\xi_\mu u^\mu,\) their velocities satisfy in the parent’s orthonormal tetrad frame
\[
\epsilon = \rho \left( E + |\vec{v}|\, |\vec{\xi}| \cos \theta \right),
\]
where \(\rho = 1/\sqrt{1-v^2}\), \(|\vec{\xi}| = \sqrt{E^2 + g_{tt}}\), and \(\theta\) is the angle between the fragment’s velocity \(\vec{v}\) and the spatial Killing vector. This leads to the Wald bounds:
\[
\rho E - \rho |\vec{v}| \sqrt{E^2 + g_{tt}} \leq \epsilon \leq \rho E + \rho |\vec{v}| \sqrt{E^2 + g_{tt}}.
\]

For the rotating Einstein--Born--Infeld (EBI) black hole with horizon radius \(r_+\) defined by \(\Delta(r_+) = 0\), the metric component at the equator simplifies to
\[
g_{tt}(r_+, \pi/2) = \frac{a^2}{r_+^2}.
\]

Using parameters \(\beta=0.1, Q=0.7, M=1, a=0.829, r_+=0.893610\), one finds
\[
g_{tt}(r_+, \pi/2) \approx 0.8606.
\]
Negative-energy fragments (\(\epsilon < 0\)) occur when their velocity is opposite the Killing vector (
\[
|\vec{v}| > \frac{E}{\sqrt{E^2 + g_{tt}(r_+)}}.
\]
For \(E=1\), this yields a threshold speed
\[
|\vec{v}| > 0.733,
\]
indicating fragments must exceed about 73.3\% of the speed of light relative to the parent to enable Penrose energy extraction~\cite{wagh1985revival}.

\subsection{Penrose Energy Extraction in Rotating Einstein--Born--Infeld Black Holes}

In the rotating spacetime of an EBI black hole, the intense frame dragging inside the ergoregion unlocks a powerful mechanism to extract rotational energy \cite{penrose1971extraction,parthasarathy1986high}. An incident particle of energy \(E^{(0)}\) entering this region may decay into two fragments: one plunging into the black hole with negative energy \(E^{(2)}<0\), while its companion escapes to infinity carrying energy \(E^{(1)}\) exceeding the original particle’s energy. This remarkable energy gain satisfies the conservation relation
\begin{equation}
E^{(0)} = E^{(1)} + E^{(2)} \quad \Rightarrow \quad E^{(1)} > E^{(0)}.
\end{equation}

To analyze this process quantitatively, define the radial velocity \(\nu = \frac{dr}{dt}\) and angular velocity \(\Omega = \frac{d\phi}{dt}\) as measured by an observer at infinity. Using the EBI metric components introduced previously, the particle’s conserved angular momentum and energy relate to the covariant time momentum \(p_t\) via
\begin{equation}
L = p_t\,\Omega, \qquad E = -p_t\,Y, \qquad Y = g_{tt} + g_{t\phi}\,\Omega, \label{a36}
\end{equation}
where the metric coefficients \(g_{\mu\nu}\) encode the nonlinear electromagnetic corrections characteristic of Born–Infeld theory  \cite{chandrasekhar1983mathematical,mtw1973gravitation}.

The four-momentum normalization,
\begin{equation}
g_{tt}\dot{t}^2 + 2 g_{t\phi} \dot{t}\dot{\phi} + g_{rr} \dot{r}^2 + g_{\phi \phi} \dot{\phi}^2 = -m^2, \label{a37}
\end{equation}
divided by \(\dot{t}^2\) and expressed using \(\Omega\) and \(\nu\), yields the key relation
\begin{equation}
g_{tt} + 2 g_{t\phi} \Omega + g_{\phi\phi} \Omega^2 + \frac{r^2 \nu^2}{\Delta_{EBI}} = -\left( \frac{m Y}{E} \right)^2. \label{a38}
\end{equation}
Since the radial velocity term \(\tfrac{r^2 \nu^2}{\Delta_{EBI}} \ge 0\), the purely angular part \(g_{tt} + 2 g_{t\phi} \Omega + g_{\phi\phi} \Omega^2\) is bounded above \cite{bardeen1972rotating}.

At the moment the particle decays within the ergosphere, conservation of energy and angular momentum impose the conditions
\begin{align}
p_t^{(0)} Y^{(0)} &= p_t^{(1)} Y^{(1)} + p_t^{(2)} Y^{(2)}, \label{40} \\
p_t^{(0)} \Omega^{(0)} &= p_t^{(1)} \Omega^{(1)} + p_t^{(2)} \Omega^{(2)}, \label{41}
\end{align}
where superscripts label the parent and resulting fragments \cite{penrose1971extraction, parthasarathy1986high}.

The efficiency of energy extraction is defined by
\begin{equation}
\eta = \frac{E^{(1)} - E^{(0)}}{E^{(0)}} = \chi - 1, \qquad \chi = \frac{E^{(1)}}{E^{(0)}} > 1. \label{a42}
\end{equation}
Combining \eqref{a36}, \eqref{40}, and \eqref{41}, one arrives at a concise expression:
\begin{equation}
\chi = \frac{(\Omega^{(0)} - \Omega^{(2)}) Y^{(1)}}{(\Omega^{(1)} - \Omega^{(2)}) Y^{(0)}}. \label{a43}
\end{equation}

To maximize energy gain, consider the initial particle with unit energy \(E^{(0)}=1\) splitting into two photons (\(m=0\)) released momentarily with zero radial velocity \(\nu=0\). The angular velocities of the fragments are chosen at the physical limits allowed in the ergoregion:
\begin{equation}
\Omega^{(1)} = \Omega_+, \qquad \Omega^{(2)} = \Omega_-, \label{a44}
\end{equation}
where \(\Omega_+\) and \(\Omega_-\) are the extremal angular velocities permitted by the rotating EBI geometry \cite{chandrasekhar1983mathematical}.

The corresponding values of \(Y\) are
\begin{equation}
Y^{(0)} = g_{tt} + g_{t\phi} \Omega^{(0)}, \qquad Y^{(1)} = g_{tt} + g_{t\phi} \Omega_+, \label{a45}
\end{equation}
where \(\Omega^{(0)}\) of the incoming null trajectory satisfies the quadratic equation derived from \eqref{a37} with \(m=0\) and \(\nu=0\):
\begin{equation}
(g_{t\phi}^2 + g_{\phi\phi}) \Omega^2 + 2 (1 + g_{tt}) g_{t\phi} \Omega + (1 + g_{tt}) g_{tt} = 0,
\end{equation}

as in the classical treatment of geodesics in stationary axisymmetric spacetimes \cite{mtw1973gravitation}

\begin{table}[htbp]
\centering
\caption{Maximum efficiency $\eta$ (\%) for different values of $a$ and $\alpha$}
\renewcommand{\arraystretch}{1.2}
\begin{tabular}{|c|c|c|c|c|c|c|c|c|c|c|}
\hline
$\beta$ & $a=0.2$ & $a=0.3$ & $a=0.4$ & $a=0.5$ & $a=0.6$ & $a=0.7$ & $a=0.8$ & $a=0.9$ & $a=1.0$ \\
\hline
0.010 & 0.27\% & 0.63\% & 1.16\% & 1.91\% & 2.95\% & 4.43\% & 6.66\% & 10.74\% & -- \\
0.020 & 0.28\% & 0.64\% & 1.19\% & 1.96\% & 3.03\% & 4.57\% & 6.93\% & 11.51\% & -- \\
0.030 & 0.28\% & 0.65\% & 1.20\% & 1.99\% & 3.08\% & 4.66\% & 7.12\% & 12.12\% & -- \\
0.040 & 0.28\% & 0.66\% & 1.22\% & 2.01\% & 3.12\% & 4.73\% & 7.27\% & 12.67\% & -- \\
0.050 & 0.29\% & 0.66\% & 1.22\% & 2.02\% & 3.15\% & 4.78\% & 7.38\% & 13.18\% & -- \\
0.100 & 0.29\% & 0.67\% & 1.25\% & 2.06\% & 3.23\% & 4.94\% & 7.75\% & 15.86\% & -- \\
0.105 & 0.29\% & 0.67\% & 1.25\% & 2.07\% & 3.23\% & 4.95\% & 7.78\% & 16.22\% & -- \\
0.110 & 0.29\% & 0.67\% & 1.25\% & 2.07\% & 3.23\% & 4.95\% & 7.80\% & 16.63\% & -- \\
0.115 & 0.29\% & 0.67\% & 1.25\% & 2.07\% & 3.24\% & 4.96\% & 7.82\% & 17.15\% & -- \\
0.120 & 0.29\% & 0.67\% & 1.25\% & 2.07\% & 3.24\% & 4.97\% & 7.84\% & 17.95\% & -- \\
0.122 & 0.29\% & 0.67\% & 1.25\% & 2.07\% & 3.24\% & 4.97\% & 7.85\% & 18.73\% & -- \\
0.123 & 0.29\% & 0.67\% & 1.25\% & 2.07\% & 3.24\% & 4.97\% & 7.86\% & -- & -- \\
0.124 & 0.29\% & 0.67\% & 1.25\% & 2.08\% & 3.24\% & 4.98\% & 7.86\% & -- & -- \\
2.000 & 0.29\% & 0.68\% & 1.26\% & 2.10\% & 3.30\% & 5.11\% & 8.32\% & -- & -- \\
5.000 & 0.29\% & 0.68\% & 1.26\% & 2.10\% & 3.30\% & 5.11\% & 8.33\% & -- & -- \\
8.000 & 0.29\% & 0.68\% & 1.26\% & 2.10\% & 3.30\% & 5.11\% & 8.33\% & -- & -- \\
\hline
\end{tabular}
\label{Tbeta}
\end{table}

\begin{table}[htbp]
\centering
\caption{Efficiency $\eta$ (\%) for EBI black hole ($\beta = 0.1$)}
\renewcommand{\arraystretch}{1.2}
\begin{tabular}{|c|c|c|c|c|c|c|c|c|c|c|}
\hline
$Q$ & $a=0.2$ & $a=0.3$ & $a=0.4$ & $a=0.5$ & $a=0.6$ & $a=0.7$ & $a=0.8$ & $a=0.9$ & $a=1.0$ \\
\hline
0.1 & 0.26\% & 0.59\% & 1.08\% & 1.77\% & 2.72\% & 4.03\% & 5.95\% & 9.11\% & -- \\
0.2 & 0.26\% & 0.60\% & 1.10\% & 1.79\% & 2.76\% & 4.09\% & 6.05\% & 9.34\% & -- \\
0.3 & 0.26\% & 0.60\% & 1.11\% & 1.83\% & 2.81\% & 4.18\% & 6.21\% & 9.68\% & -- \\
0.4 & 0.27\% & 0.62\% & 1.14\% & 1.87\% & 2.87\% & 4.29\% & 6.41\% & 10.14\% & -- \\
0.5 & 0.27\% & 0.63\% & 1.16\% & 1.91\% & 2.95\% & 4.43\% & 6.66\% & 10.74\% & -- \\
0.6 & 0.28\% & 0.65\% & 1.20\% & 1.97\% & 3.05\% & 4.59\% & 6.96\% & 11.54\% & -- \\
0.7 & 0.29\% & 0.67\% & 1.23\% & 2.03\% & 3.16\% & 4.78\% & 7.33\% & 12.64\% & -- \\
0.8 & 0.30\% & 0.69\% & 1.28\% & 2.11\% & 3.29\% & 5.01\% & 7.77\% & 14.30\% & -- \\
0.9 & 0.31\% & 0.71\% & 1.32\% & 2.20\% & 3.44\% & 5.27\% & 8.32\% & 17.89\% & -- \\
1.0 & 0.32\% & 0.74\% & 1.38\% & 2.29\% & 3.61\% & 5.58\% & 9.00\% & -- & -- \\
\hline
\end{tabular}
\label{TQ}
\end{table}

with the physically relevant root
\begin{equation}
\Omega^{(0)} = \frac{-(1 + g_{tt}) g_{t\phi} + \sqrt{(1 + g_{tt})(g_{t\phi}^2 - g_{tt} g_{\phi\phi})}}{g_{t\phi}^2 + g_{\phi\phi}}.
\end{equation}\cite{chandrasekhar1983mathematical, mtw1973gravitation}.

Substituting \eqref{a44}–\eqref{a45} into \eqref{a43} yields the local efficiency at the decay point:
\begin{equation}
\eta = \frac{(g_{tt} + g_{t\phi} \Omega_+) (\Omega^{(0)} - \Omega_-)}{(g_{tt} + g_{t\phi} \Omega^{(0)})(\Omega_+ - \Omega_-)} - 1.
\end{equation} \cite{penrose1971extraction, parthasarathy1986high}.

The absolute maximum of \(\eta\) occurs as the decay point approaches the outer event horizon \(r = r_+\), defined by \(\Delta_{EBI}(r_+) = 0\) \cite{bardeen1972rotating}. Expressing the EBI black hole’s effective charge squared at the horizon as \({Q_{eff}}^2(r_+; a, \beta, Q)\), which encodes the nonlinear Born–Infeld corrections dependent on spin \(a\), the Born–Infeld parameter \(\beta\), and charge \(Q\), the maximal efficiency takes the elegant closed form
\begin{equation}
\eta_{\max}^{\mathrm{EBI}} = \frac{1}{2} \left( \sqrt{\frac{2 M r_+ - {Q_{eff}}^2 (r_+; a, \beta, Q)}{r_+^{2}}} - 1 \right).
\end{equation} \cite{garcia1999einstein}.
This smoothly recovers the classical Kerr–Newman result as \(\beta \to \infty\) (the linear electrodynamics limit) \cite{chandrasekhar1998mathematical}.

This expression reveals the subtle influence of nonlinear electromagnetic fields on the maximal extractable energy from rotating black holes. The Born–Infeld parameter \(\beta\) effectively modifies the distribution of electromagnetic energy, reshaping the ergoregion and consequently the upper bound on the Penrose process efficiency. As a result, EBI black holes furnish an intriguing arena where classical black hole thermodynamics meets nonlinear gauge field dynamics, with potential astrophysical implications.

\begin{figure}[htbp]
    \centering

    \makebox[\textwidth][c]{
        \includegraphics[width=0.55\textwidth]{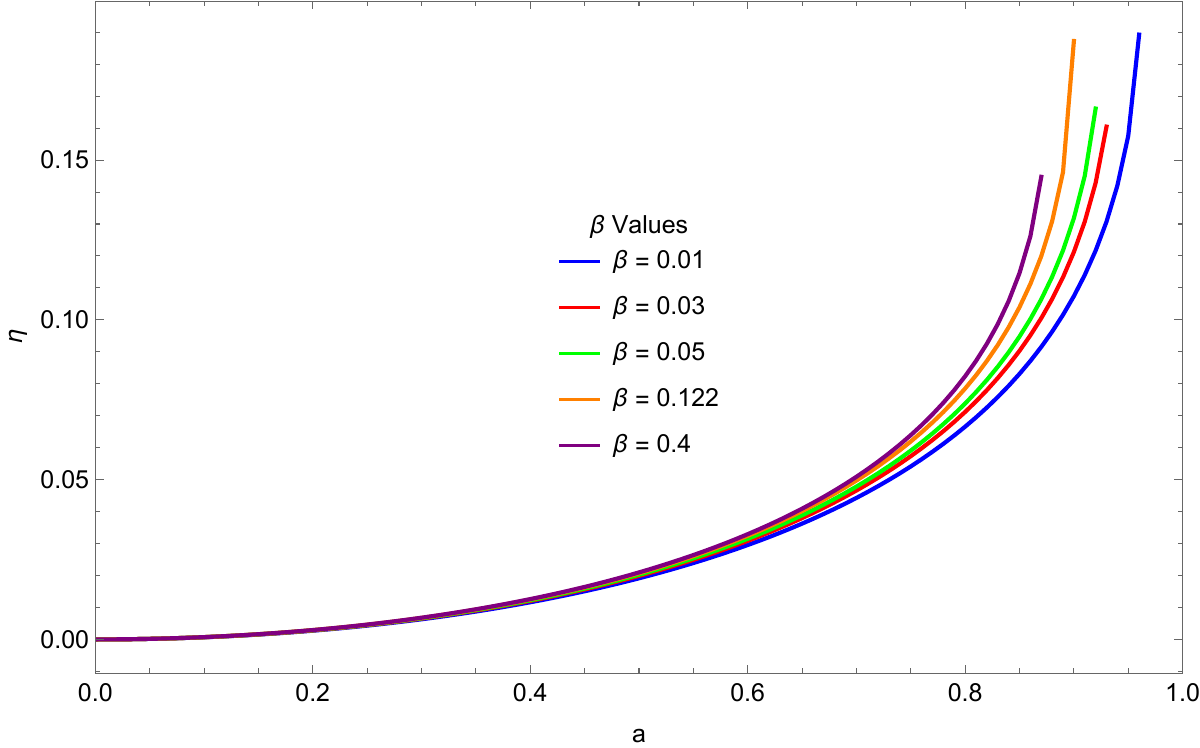}
        \hspace{2em}
        \includegraphics[width=0.55\textwidth]{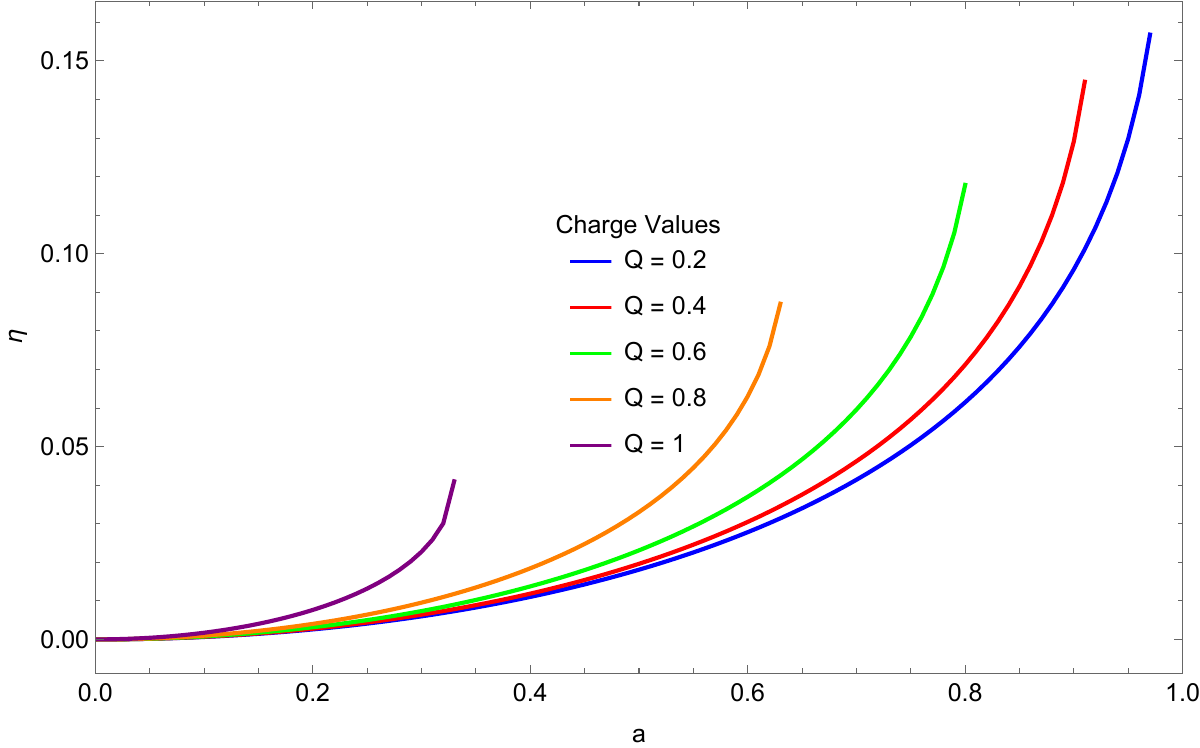}
    }

    \caption{Efficiency \(\eta\) versus spin parameter \(a\) for Einstein--Born--Infeld black holes.}
    \label{fig:Efficiency}
\end{figure}

The plots in Fig. \ref{fig:Efficiency} and Tables \ref{Tbeta} and \ref{TQ}  illustrate the efficiency $\eta$ of the EBI black hole as a function of the spin parameter $a$ for various values of the $\beta$ and the charge $Q$. The results demonstrate that efficiency $\eta$ increases monotonically with increasing spin $a$ for all cases. For fixed $Q$, larger values of $\beta$ slightly enhance the efficiency, especially at higher spin. Similarly, for fixed $\beta$, increasing the charge $Q$ leads to a noticeable rise in $\eta$, with the largest differences appearing at greater spin. In summary, both higher $\beta$ and larger $Q$ yield greater efficiency for the EBI black hole, and the effect of these parameters becomes more significant as the spin approaches its maximum value.

\subsection{Comparative analysis}

\begin{table}[htbp]
\centering
\renewcommand{\arraystretch}{1.2}
\begin{tabular}{|c|c|c|c|}
\hline
\textbf{a} & \textbf{Kerr $\eta$ (\%) (Q = 0)} & \textbf{Kerr--Newman $\eta$ (\%) (Q = 0.5)} & \textbf{EBI $\eta$ (\%) (Q = 0.5, $\beta=0.1$)} \\
\hline
0.2   & 0.25\%  & 0.29\%  & 0.29\%  \\ \hline
0.3   & 0.59\%  & 0.68\%  & 0.67\%  \\ \hline
0.4   & 1.08\%  & 1.26\%  & 1.25\%  \\ \hline
0.5   & 1.76\%  & 2.10\%  & 2.06\%  \\ \hline
0.6   & 2.70\%  & 3.30\%  & 3.23\%  \\ \hline
0.7   & 4.01\%  & 5.11\%  & 4.94\%  \\ \hline
0.8   & 5.90\%  & 8.33\%  & 7.75\%  \\ \hline
0.85  & 7.23\%  & 11.88\% & 10.19\% \\ \hline
0.866 & 7.73\%  & 15.96\% & 11.32\% \\ \hline
0.9   & 9.01\%  & --      & 15.86\% \\ \hline
1.0   & 20.71\% & --      & --      \\ \hline
\end{tabular}
\caption{Comparison of energy extraction efficiency $\eta$ for Kerr, Kerr--Newman, and EBI black holes with $Q=0.5$ and $\beta=0.1$.}
\label{tab:efficiency}
\end{table}

The comparative analysis of energy extraction efficiencies for Kerr, Kerr-Newman, and EBI black holes reveals clear and physically significant trends. As established in the literature, Kerr black holes ($Q=0$) achieve the highest theoretical efficiency, reaching approximately 20.7\% in the extremal spin limit, due to the absence of charge and purely gravitational energy extraction mechanisms~\cite{Penrose1969, Bardeen1972, Wald1984}. Introducing electric charge, as in the Kerr-Newman case ($Q=0.5$), reduces the maximum efficiency for all spin parameters; this suppression is attributed to the effect of electromagnetic fields on the ergoregion and horizon structure~\cite{Christodoulou1972, Carter1968}. When Born–Infeld nonlinear electrodynamics is incorporated (EBI black holes), efficiencies become systematically lower than for Kerr-Newman at the same spin and charge. Our results, covering a broad range of spin parameters and Born–Infeld coupling $\beta$, demonstrate that efficiency in EBI black holes increases smoothly with $\beta$, approaching the Kerr-Newman value as nonlinearity weakens. For strong nonlinearity ($\beta=0.1$), the efficiency is notably lower—for example, at $a=0.9$, $\eta$ rises from 15.86\% at $\beta=0.1$ to 18.73\% at $\beta=0.122$. This ordering, $\eta_\text{Kerr} > \eta_\text{Kerr-Newman} > \eta_\text{EBI}$, directly reflects the progressive reduction in extractable energy due to both electric charge and nonlinear electromagnetic corrections. As $\beta$ increases, the EBI efficiency converges to its Kerr-Newman counterpart, thus elucidating the physical impact of nonlinear electrodynamics on black hole energetics and demonstrating the critical role of both charge and field nonlinearity in limiting the Penrose process~\cite{Penrose1969, Bardeen1972, Christodoulou1972}.

\section{Irreducible Mass: Geometric Proxy for Black Hole Entropy} \label{S7}

The irreducible mass, \( M_{\text{irr}} \), represents the portion of a rotating black hole's mass that cannot be extracted by classical processes, thereby setting a fundamental upper limit on the extractable rotational energy \cite{christodoulou1970reversible}. In the context of the EBI black hole, the nonlinear electromagnetic parameter \(\beta\) and the electric charge \(Q\) modify the geometry of the spacetime, particularly the event horizon structure, which in turn influences \( M_{\text{irr}} \).

The outer event horizon radius \( r_+ \) is the largest real root of the equation $\Delta_{\mathrm{EBI}}(r) = 0$. The nonlinear Born–Infeld corrections and electric charge enter implicitly through the horizon radius \(r_+\). The irreducible mass is connected to the horizon area \(A\) \cite{christodoulou1971} by
\begin{equation}
M_{\text{irr}} = \sqrt{\frac{A}{16\pi}},
\end{equation}
with the horizon area computed via
\begin{equation}
A = \int_0^{2\pi} \! d\phi \int_0^\pi \! d\theta \, \sqrt{g_{\theta\theta} g_{\phi\phi}} \Big|_{r = r_+} = 4 \pi (r_+^2 + a^2).
\end{equation}
So the irreducible mass takes the form
\begin{equation}
M_{\text{irr}} = \frac{1}{2} \sqrt{r_+^2 + a^2},
\end{equation}
Since the black hole entropy \(S\) in semiclassical gravity is proportional to the horizon area, the irreducible mass \(M_{\text{irr}}\) acts as a direct thermodynamic proxy reflecting how nonlinear electromagnetic effects impact black hole entropy.

Tables \ref{TbetaMirr} and \ref{TQMirr} illustrate numerical values of the horizon radius \(r_+\) and irreducible mass \(M_{\text{irr}}\) for varying Born–Infeld parameter \(\beta\), electric charge \(Q\), and spin parameter \(a\). In particular, Table \ref{TbetaMirr} shows how \(r_+\) and \(M_{\text{irr}}\) evolve with \(\beta\) for fixed \(Q\) and different spins, while Table \ref{TQMirr} displays their dependence on \(Q\) for fixed \(\beta\).

\begin{table}[htbp]
\centering
\caption{Event horizon radius \(r_+\) and irreducible mass \(M_{\text{irr}}\) for different values of \(\beta\) and spin parameter \(a\).}
\renewcommand{\arraystretch}{1.2}
\begin{tabular}{|c|c|c|c|c|c|c|}
\hline
\(\beta\) & \multicolumn{2}{c|}{\(a = 0.2\)} & \multicolumn{2}{c|}{\(a = 0.3\)} & \multicolumn{2}{c|}{\(a = 0.5\)} \\
\hline
          & \(r_+\) & \(M_{\text{irr}}\) & \(r_+\) & \(M_{\text{irr}}\) & \(r_+\) & \(M_{\text{irr}}\) \\
\hline
0.01 & 1.91031 & 0.960374 & 1.8832 & 0.953475 & 1.7905 & 0.929502 \\
0.03 & 1.88000 & 0.945306 & 1.8520 & 0.938071 & 1.75584 & 0.912822 \\
0.05 & 1.86675 & 0.938715 & 1.83819 & 0.931256 & 1.73988 & 0.905149 \\
0.07 & 1.85936 & 0.935041 & 1.83042 & 0.927422 & 1.73059 & 0.900685 \\
0.10 & 1.85316 & 0.931961 & 1.82384 & 0.924175 & 1.72244 & 0.896774 \\
0.30 & 1.84429 & 0.927549 & 1.81425 & 0.919442 & 1.70975 & 0.890679 \\
0.50 & 1.84324 & 0.927031 & 1.81310 & 0.918875 & 1.70811 & 0.889894 \\
0.70 & 1.84294 & 0.92688  & 1.81276 & 0.918709 & 1.70763 & 0.889662 \\
1.00 & 1.84277 & 0.926798 & 1.81258 & 0.91862  & 1.70736 & 0.889536 \\
3.00 & 1.84263 & 0.926782 & 1.81242 & 0.918542 & 1.70714 & 0.889426 \\
5.00 & 1.84262 & 0.926722 & 1.81241 & 0.918536 & 1.70712 & 0.889417 \\
7.00 & 1.84262 & 0.92672  & 1.81241 & 0.918534 & 1.70711 & 0.889414 \\
9.00 & 1.84262 & 0.92672  & 1.81241 & 0.918534 & 1.70711 & 0.889413 \\
\hline
\end{tabular}
\label{TbetaMirr}
\end{table}

\begin{table}[htbp]
\centering
\caption{Event horizon radius \(r_+\) and irreducible mass \(M_{\text{irr}}\) for different values of \(Q\) and spin parameter \(a\).}
\renewcommand{\arraystretch}{1.2}
\begin{tabular}{|c|c|c|c|c|c|c|}
\hline
\(Q\) & \multicolumn{2}{c|}{\(a = 0.2\)} & \multicolumn{2}{c|}{\(a = 0.3\)} & \multicolumn{2}{c|}{\(a = 0.5\)} \\
\hline
       & \(r_+\) & \(M_{\text{irr}}\) & \(r_+\) & \(M_{\text{irr}}\) & \(r_+\) & \(M_{\text{irr}}\) \\
\hline
0.1 & 1.97543 & 0.992764 & 1.94948 & 0.986212 & 1.8612  & 0.963594 \\
0.2 & 1.96491 & 0.987534 & 1.93876 & 0.980919 & 1.84973 & 0.958059 \\
0.3 & 1.95089 & 0.980161 & 1.92368 & 0.973466 & 1.83364 & 0.950296 \\
0.4 & 1.93174 & 0.971033 & 1.90501 & 0.964243 & 1.81375 & 0.940701 \\
0.5 & 1.91031 & 0.960374 & 1.88320 & 0.953475 & 1.7905  & 0.929502 \\
0.6 & 1.88609 & 0.948334 & 1.85856 & 0.941311 & 1.76421 & 0.916848 \\
0.7 & 1.85931 & 0.93502  & 1.83131 & 0.927859 & 1.73508 & 0.902843 \\
0.8 & 1.83013 & 0.920514 & 1.8016  & 0.913201 & 1.70326 & 0.887566 \\
0.9 & 1.79868 & 0.904884 & 1.76955 & 0.897402 & 1.66886 & 0.871075 \\
1.0 & 1.76507 & 0.888181 & 1.73529 & 0.880514 & 1.63195 & 0.853414 \\
\hline
\end{tabular}
\label{TQMirr}
\end{table}

Several notable trends emerge from these numerical results. As the Born–Infeld parameter \(\beta\) increases, nonlinear electromagnetic effects weaken, causing the irreducible mass \(M_{\text{irr}}\) to approach values closer to the classical Kerr–Newman limit~\cite{cheng2025rotating}. Conversely, smaller \(\beta\), corresponding to stronger nonlinear effects, decreases \(M_{\text{irr}}\), indicating a smaller horizon and reduced entropy. Increasing the electric charge \(Q\) consistently lowers both the horizon radius and \(M_{\text{irr}}\) across all spin values, reflecting electromagnetic repulsion that shrinks the horizon area and lowers black hole entropy. This effect is more pronounced at higher spin parameters \(a\), highlighting the complex interplay between rotation and charge in shaping the thermodynamic configuration of the EBI black hole.

The extractable energy, given by
\begin{equation}
E_{\text{ext}} = M - M_{\text{irr}},
\end{equation}
thus grows as \(M_{\text{irr}}\) decreases, implying enhanced efficiency in energy extraction mechanisms such as the Penrose process, particularly in regimes dominated by Born–Infeld nonlinearity and higher charge configurations \cite{christodoulou1971}.

These results highlight the significant role of nonlinear electromagnetic dynamics encoded in the Born–Infeld parameter. Measuring variations in irreducible mass and horizon metrics provides a quantitative probe into how nonlinear electrodynamics modifies entropy bounds and constrains reversible energy extraction from rotating EBI black holes \cite{bardeen1973,Dey2004BornInfeldBH}.

 \section{Conclusion} \label{S8}

In this work, we have conducted a comprehensive analysis of rotational energy extraction from rotating EBI black holes, incorporating the effects of nonlinear Born--Infeld electrodynamics on the black hole geometry and ergoregion structure. By deriving particle kinematics and evaluating the Penrose process within the equatorial plane, we obtained explicit expressions for the maximal energy extraction efficiency as a function of the black hole spin, electric charge, and the Born--Infeld parameter \( \beta \).

Our numerical investigations reveal that increasing the electric charge and strengthening the nonlinear Born--Infeld corrections reduce both the event horizon radius and the size of the ergoregion. These geometric modifications generally suppress the Penrose process efficiency compared to classical Kerr and Kerr--Newman black holes. However, at certain spins and values of \( \beta \), the EBI geometry modifies the horizon and ergoregion structures in such a way that the efficiency can locally exceed that of Kerr--Newman. We analyzed the angular velocity distributions and conditions for negative energy states within the ergoregion, which underpin the Penrose process mechanism and influence energy extraction limits.

Furthermore, we examined the irreducible mass, a geometric quantity proportional to the horizon area and serving as a proxy for black hole entropy. We demonstrated that stronger nonlinear Born--Infeld electrodynamics and higher charge decrease the irreducible mass and horizon area, implying an increased fraction of the black hole mass is potentially extractable as rotational energy under favorable parameters.

Our results highlight the crucial influence of nonlinear electromagnetic dynamics on the causal structure and energy extraction processes of rotating black holes. By linking geometric modifications, particle motion, and the irreducible mass as an entropy-related proxy, this study enhances the understanding of energy extraction mechanisms in nonlinear electrodynamic spacetimes. These findings carry implications for astrophysical scenarios with strong electromagnetic fields. Future work may extend this analysis to charged particles, alternative spacetime asymptotics, and observational signatures specific to Einstein--Born--Infeld black holes.

\end{document}